\documentclass[onecolumn,showpacs,aps,epsfig,nofootinbib]{revtex4}
\usepackage{amssymb, amsmath, bm, dcolumn, epsf, graphicx, latexsym, mathbbol, slashed}
\usepackage{graphicx}
\usepackage{dcolumn}
\usepackage{bm}
\usepackage{amssymb}
\usepackage{color}
\usepackage{multirow}
\usepackage[colorlinks,linkcolor=blue,citecolor=blue,urlcolor=blue]{hyperref}

 \def\tskip{\setlength{\tskip}{5pt}}
\def\colwidth{\setlength{\colwidth}{3.5in}}

\newcommand{\lsim}{\mathrel{\hbox{\rlap{\lower.55ex\hbox{$\sim$}} \kern-.3em \raise.4ex \hbox{$<$}}}}
\newcommand{\gsim}{\mathrel{\hbox{\rlap{\lower.55ex\hbox{$\sim$}} \kern-.3em \raise.4ex \hbox{$>$}}}}
\newcommand{\beq}{\begin{equation}}
\newcommand{\eeq}{\end{equation}}
\newcommand{\be}{\begin{equation}}
\newcommand{\ee}{\end{equation}}
\newcommand{\bes}{\begin{equation*}}
\newcommand{\ees}{\end{equation*}}
\newcommand{\beqa}{\begin{eqnarray}}
\newcommand{\eeqa}{\end{eqnarray}}
\newcommand{\bea}{\begin{eqnarray}}
\newcommand{\ena}{\end{eqnarray}}

\usepackage{color}
\def\b {\color{blue}}
\def\r {\color{black}}
\def \mSun {${\rm M}_\odot$}
\begin{document}

\title{Localization accuracy of compact binary coalescences detected by the third-generation gravitational-wave detectors and implication for cosmology}

\author{Wen Zhao}
\email[]{wzhao7@ustc.edu.cn}
\affiliation{CAS Key Laboratory for Researches in Galaxies and Cosmology, Department of Astronomy, University of Science and Technology of China, Chinese Academy of Sciences, Hefei, Anhui 230026, China \\
School of Astronomy and Space Science, University of Science and Technology of China, Hefei 230026, China}

\author{Linqing Wen }
\email[]{linqing.wen@uwa.edu.au}
\affiliation{School of Physics, The University of Western Australia, 35 Stirling Hwy, Crawley, WA 6009, Australia}

\date{\today}

%{\bf 1. Why $\alpha\sim O(1)$? 2. Galactic matter density or galactic matter density? 3. P12, left column, [?] and [?]?; 4. change pdf figure to eps figure}

\begin{abstract}
{
We use the Fisher information matrix to investigate the angular resolution and luminosity distance uncertainty for coalescing binary neutron stars (BNSs) and neutron star-black hole binaries (NSBHs) detected by the third-generation (3G) gravitational-wave (GW) detectors. Our study focuses on an individual 3G detector and a network of up to four 3G detectors at different locations including the US, Europe, China and Australia for the proposed Einstein Telescope (ET) and Cosmic Explorer (CE) detectors. We in particular examine the effect of the Earth's rotation, as GW signals from BNS and low mass NSBH systems could be hours long for 3G detectors. In this case, an individual detector can be effectively treated as a detector network with long baselines formed by the trajectory of the detector as it rotates with the Earth. Therefore, a single detector or two-detector networks could also be used to localize the GW sources effectively. We find that, a time-dependent antenna beam-pattern function can help better localize BNS and NSBH sources, especially those edge-on ones. The medium angular resolution for one ET-D detector is around 150 deg$^2$ for BNSs at a redshift of $z=0.1$, which improves rapidly with a decreasing low-frequency cutoff $f_{\rm low}$ in sensitivity. The medium angular resolution for a network of two CE detectors in the US and Europe respectively is around 20 deg$^2$ at $z=0.2$ for the simulated BNS and NSBH samples. While for a network of two ET-D detectors, the similar angular resolution can be achieved at a much higher redshift of $z=0.5$. The angular resolution of a network of three detectors is mainly determined by the baselines between detectors regardless of the CE or ET detector type.  The medium angular resolution of BNS for a network of three detectors of the ET-D or CE type in the US, Europe and Australia is around 10 deg$^2$ at $z=2$. We discuss the implications of our results to multi-messenger astronomy and in particular to using GW sources as independent tools to constrain the Hubble constant $H_0$, the deceleration parameter $q_0$ and the equation-of-state (EoS) of dark energy. We find that in general, if 10 BNSs or NSBHs at $z=0.1$ with known redshifts are detected by 3G networks consisting of two ET-like detectors, $H_0$ can be measured with an accuracy of $0.9\%$. If 1000 face-on BNSs at $z<2$ are detected with known redshifts, we are able to achieve $\Delta q_0=0.002$ for deceleration parameter, or $\Delta w_0=0.03$ and $\Delta w_a=0.2$ for EoS of dark energy, respectively.
}
\end{abstract}


\maketitle

\section{Introduction \label{section1}}

%$^{\circ}$
%{\r [Add more test of GR and cosmology papers from the LVC, add GW170104 since already announced]}

The first detection of the gravitational-wave (GW) event GW150914 \cite{gw150914,150914-localization,150914-testGR}, as well as the events GW151226 \cite{gw151226}, GW170104 \cite{gw170104}, GW170608 \cite{gw170608}, GW170814 \cite{gw170814}, GW170817 \cite{gw170817} and the less significant candidate LVT151012 \cite{151012}, marks the beginning of the era of GW astronomy. Since GWs are expected to be produced in extreme conditions, including the strong gravitational fields, high density regions, and/or extremely early stage of the Universe, and propagate nearly freely in the spacetime once generated, they encode the clean information of these extreme conditions.   Currently, the second-generation (2G) ground-based laser interferometer GW detectors are either ongoing and successfully making discoveries of GWs or are expected to be operational in the next few years.  This includes Advanced LIGO \cite{Advanced LIGO}, Advanced Virgo \cite{Advanced Virgo}, KAGRA in Japan \cite{KAGRA}, and the proposed LIGO-India \cite{LIGO-India}.

\begin{figure}
\begin{center}
\centerline{\includegraphics[width=10cm]{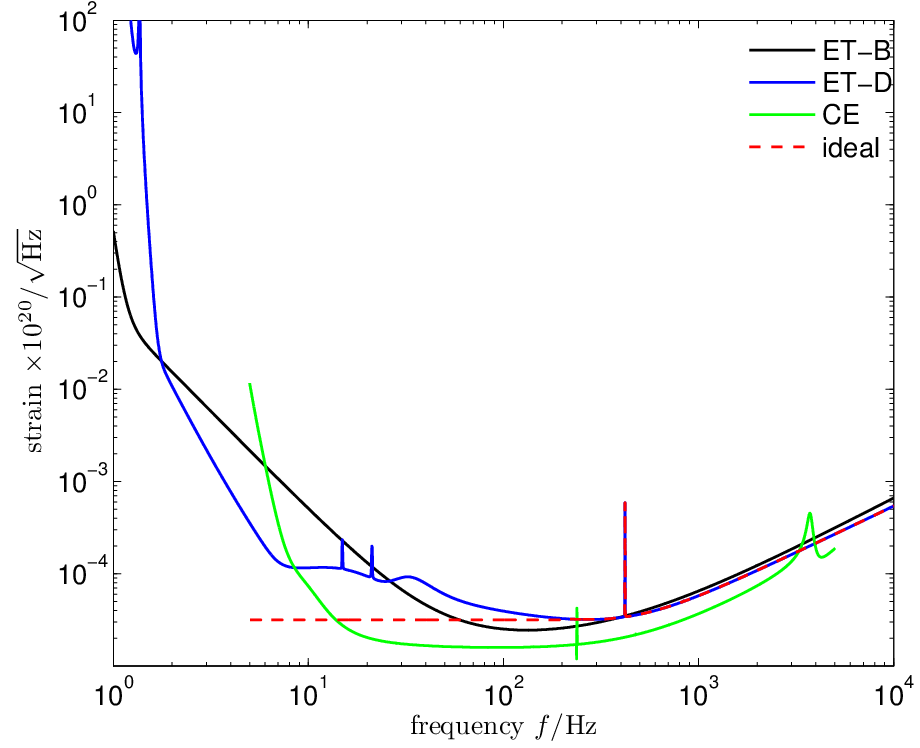}}
\end{center}\caption{The amplitude spectral density of ET with ET-B noise level (solid black line) \cite{ET2}, ET with ET-D noise level (solid blue line) \cite{ET2}, CE (solid green line) \cite{ce}, and the ideal experiment (dashed red line). }\label{f0}
\end{figure}

Looking forward, two leading proposals are currently under consideration for the design of the third-generation (3G) GW detectors (see Fig.~\ref{f0} for their sensitivity designs). One is the Einstein Telescope (ET), which is a proposed European GW observatory \cite{et,etn,ET2}.  The other is Cosmic Explorer (CE) \cite{ce,CE2}, which is a proposed US-based future GW detector.  ET consists of three Michelson interferometers with 10 km long arms, and inter-arm angle of $60^{\circ}$, arranged to form an equilateral triangle.  Two sensitivity estimates have been put forward for the ET.  One is based on a single interferometer covering the full frequency range from 1 Hz to 10 kHz, and is referred to as ET-B (Fig.~\ref{f0}, black solid line) \cite{ET-B}. The other one uses the so-called xylophone design \cite{ET-D}, in which one GW detector is composed of one cryogenic low-frequency interferometer and one room temperature high-frequency interferometer.  This sensitivity model is referred to as ET-D (Fig.~\ref{f0}, blue solid line). Relative to ET-B, the noise in sub-10 Hz band of ET-D is substantially reduced. The scientific potentials of ET have been studied by many authors \cite{sathya2009,zhao-2011,gair2,cai,zhu,mock,gair,et-science,d1,d2,d3,vilta,zhao-2017,et-bd,zhu2017}. Different from the ET designs, CE (Fig.~\ref{f0}, green solid line) will keep L-shaped configuration, but the length of arms will be significantly increased to 40 km \cite{ce}.  In this design, the lower limit of sensitive band is determined by the seismic and Newtonian noise, the latter of which is still poorly understood \cite{ce}.  Compared to ET-D, the target sensitivity of CE is significantly better at high frequencies above $\sim 8$ Hz but ET-D prevails at lower frequencies at around 1-8 Hz. With CE, the horizon redshift of detectable binary black holes (BBHs) will extend to $z\sim 7$, and that of binary neutron stars (BNSs) will be about 2 \cite{CE2}.

The prime candidate sources for both 2G and 3G detectors are GWs produced by coalescing binary systems of compact objects: BNSs, neutron star-black hole (NSBH) binaries, and the BBHs \cite{gw-waveform}. With more binary coalescence sources expected to be detected with high SNR and at greater distances, it is of paramount interest to use these sources as an independent tool to study the evolution history of the Universe \cite{schutz1986,GW-cos1,GW-cos2,GW-cos3,GW-cos4,GW-cos5,GW-cos6,sathya2009,zhao-2011}, and to test gravity in the strong gravitational fields \cite{testGR,150914-testGR,testGR1,testGR2,testGR3,testGR4,testGR5,testGR6,testGR7,testGR8,testGR9,zhao-2017,gw170814,gw1708142}. % Thus, a great deal of effort has been devoted to the research and detection of GWs. %For them, the localization of sources, including the angular resolution and the luminosity distance error, by GW observations plays a crucial role for GW astronomy.
Accurate localization of these GW sources is crucial to help follow-up observations of their electromagnetic (EM) or neutrino counterparts \cite{150914-localization,em-counterpart}, and to increase the chances of identifying their host galaxies and to determine their redshifts. On the other hand, the angular and distance resolutions of an individual 2G detector are rather poor \cite{thorne1987}.  GW detector network with large baselines is therefore needed to facilitate better source localization \cite{schutz-network,vilta}. However, the tens of square degree nominal angular resolution of 2G detectors still pose a challenge for effective follow-up observations by conventional EM telescopes \cite{ai,a0,a1,a2,network1,network2,algorithm1,algorithm2,algorithm3,algorithm4,algorithm5,algorithm6}. A network of multiple detectors is also necessary to help measure both GW polarizations, which in turn helps improve the estimation of the source luminosity distance. In addition, a detector network can also help measure the extra polarization modes of GW allowed in more general theories of gravitation, which is of crucial importance to test gravity in strong gravitational fields \cite{will-book,WenSchutz2005,yunes2012}.
% Relative to an individual detector, the network consisting of several GW detectors can greatly improve the localization of GW sources \cite{schutz-network,et-ce}.

{
One key difference between the proposed 3G and 2G detectors is that the proposed 3G detectors are expected to have extended sensitivity at lower frequencies than 2G detectors.    For ET-D, the cutoff of the low-frequency sensitivity is extended to $f_{\rm low} \sim 1$ Hz.   The time to coalescence of a BNS is a function of $f^{-8/3}_{\rm low}$.} The BNS signals can therefore be in band for 3G detectors for several hours, or even for several days compared to tens of minutes in the 2G detector band.  Thus, the time dependence of the detector beam-pattern functions of the 3G detectors, due to the motion of the Earth, could become important. With this effect, an individual detector can be effectively treated as a network including a set of detectors at different locations along the detector's trajectory on the Earth, which observe a given GW event at different time. The baseline of this `network' is determined by the duration of the event in the detector's band. If the duration can be around one day, the baseline is roughly the Earth's diameter.  We therefore anticipate a significant improvement in parameter estimation, especially that related to the localization for detectors with significant low-frequency sensitivity \footnote{We should emphasize that, in addition to the loss of the localization, the bias of the parameter estimation might also be induced, if ignoring the time dependence of the detector antenna beam-pattern functions. However, in order to address this issue, the Monte Carlo analyzes instead of the Fisher matrix analyzes are needed, which is beyond the scope of the present work.}.%  Note the low-frequency cutoff of 2G detectors is around 10 Hz at design sensitivity.  The GW signals of compact binaries observed by these detectors, can be in band for much less than one hour.  The antenna beam-pattern functions can therefore be treated as constant for a given GW event.

%%%%%%%%%%%%%%%%%%%%%%%%%%%%%%%%%%%%%%%%%  figure 1, figure 1, figure1%%%%%%%%%%%%%%%%%%%%%%%%%%%%%%%%%%%%%%%%%%%%%%%%%%
%%%%%%%%%%%%%%%%%%%%%%%%%%%%%%%%%%%%%%%%  figure 1, figure 1, figure1%%%%%%%%%%%%%%%%%%%%%%%%%%%%%%%%%%%%%%%%%%%%%%%%%%

% In this article, we will consider both designs of ET. The target sensitivity for ET is more than one order improvement in comparison to current advanced detectors.

Several algorithms, including numerical computations and analytical expressions, have been developed to quantify the effectiveness of GW detector networks \cite{ai,a0,a1,a2,network1,network2,algorithm1,algorithm2,algorithm3,algorithm4,algorithm5}. In this article, we use the standard Fisher matrix technique to investigate the angular resolution and luminosity distance determination of a single 3G detector as well as a network of multiple detectors including ET, CE, ET-like and/or CE-like detectors at different locations \footnote{In finishing this paper, we find a parallel independent work \cite{fairhurst2017} that has some overlap with our calculation of the angular resolution for a network of more than two 3G detectors.}. We also demonstrate the potential capabilities of utilizing the accurate localization of GW sources to constrain the cosmological parameters, including the Hubble constant, the deceleration parameter and the equation-of-state (EoS) of cosmic dark energy. {In addition, to study the effect of low-frequency sensitivity, we also include an ideal detector (Fig.~\ref{f0}, red dashed line) with a better low frequency sensitivity than ET-D.  The detector is designed to have a constant sensitivity across the frequency band of $f\in[5,400]$ Hz} {with the noise power spectral density $S_I(f)=10^{-49}/{\rm Hz}$} and infinity at $f< 5$ Hz.

{
This paper is organized as follows.  In Sec.~\ref{sec:GW}, we review the mathematical preliminaries of GW waveforms from binary coalescence and the response of GW detectors to these signals.  In Sec.~\ref{sec:Fisher}, we describe our method to calculate the Fisher matrix for a set of nine parameters required to fully describe a GW signal from compact binary coalescence.  A comparison with a previously published result is presented to verify the validity of our calculation.  In Sec.~\ref{sec:One3G} we show the result for one 3G detector from numerical simulations for both BNS and NSBH sources. We demonstrate the dependence of the angular resolution and distance determination on the low-frequency cutoff of the detector.  In Sec.~\ref{sec:Network}, we present the localization of GW sources at different redshifts observed by an individual 3G detector, and by a network of two to four 3G detectors. In Sec.~\ref{sec:Cosmology}, we discuss the implications for cosmology in the 3G detector era. At the end, in Sec.~\ref{sec:Concl}, we summarize our main results.
}

Throughout this paper, we choose the units in which $G=c=1$, where $G$ is the Newtonian gravitational constant, and $c$ is the speed of light in vacuum.

\section{GW waveforms and the detector response\label{section2}}
\label{sec:GW}
As a general consideration, we assume that a GW event is observed by a ground-based network, which includes $N_d$ GW detectors, each with spatial size much smaller than the GW wavelength \footnote{In the recent paper \cite{essick2017}, the authors discussed the effects when the GW frequencies are comparable to the round-trip light travel time down the detector arms.}. The spatial locations of the detectors are given by the vector ${\rm {\bf r}}_I$ with $I=1,2,...,N_d$. In the celestial coordinate system, the vector ${\rm {\bf r}}_I$ is given by
\begin{equation}
{\rm {\bf r}}_I = R_{\oplus}(\sin\varphi_I\cos\alpha_I,~\sin\varphi_I\sin\alpha_I,~\cos\varphi_I),
\end{equation}
where $R_{\oplus}$ is the radius of the Earth, $\varphi_I$ is the latitude of the detector. The angle $\alpha_I$ is defined as $\alpha_I\equiv \lambda_I+\Omega_r t$, where $\lambda_I$ is the east longitude of the detector, $\Omega_r$ is the rotational angular velocity of the Earth. Note that, throughout this paper we fix $t=0$ at which the Greenwich sidereal time is zero.

For an individual detector labeled by $I$, the response to an incoming GW signal is a linear combination of two wave polarizations in the transverse-traceless gauge,
\begin{equation}
d_I(t_0+\tau_I+t)=F_I^{+} h_+(t)+F_I^{\times}h_{\times}(t),~~~~0<t<T,
\end{equation}
where $h_+$ and $h_{\times}$ are the plus and cross modes of GW respectively, $t_0$ is the arrival time of the wave at the coordinate origin and $\tau_I$ is the time required for the wave to travel from the origin to reach the $I$-th detector at time $t$,
\begin{equation}\label{tau_I}
\tau_I(t)={\rm {\bf n}}\cdot{\rm {\bf r}}_I(t).
\end{equation}
Here ${\rm {\bf n}}$ is the propagation direction of a GW,
$t\in[0,T]$ is the time label of the wave, and $T$ is the signal
duration. The quantities $F_I^{+}$ and $F_I^{\times}$ are the
detector's antenna beam-pattern functions, which depend on the source location $(\theta_s,\phi_s)$, the polarization angle $\psi_s$, the detector's location on the Earth labeled by latitude $\varphi$, longitude $\lambda$, the angle $\gamma$, which determines the orientation of the detector's arms with respect to local geographical directions: $\gamma$ is measured counter-clockwise from East to the bisector of the interferometer arms, as well as the angle between the interferometer arms $\zeta$. See in Table \ref{table1} \cite{vilta} for the parameters used for the potential Einstein Telescope (ET) in Europe, Cosmic Explorer (CE) experiment in the US, the assumed detector in Australia, and that in China, respectively. For the ground-based detectors, the expressions of $F_I^{+}$ and $F_I^{\times}$ are explicitly given in \cite{schutz1998}, which are periodic functions of time with a period equal to one sidereal day, due to the diurnal motion of the Earth.

During the inspiral phase of the binary coalescence, the change in orbital frequency over a single GW cycle is
negligible, it is therefore possible to apply a stationary phase
approximation (SPA) to compute the Fourier transformation. We denote the Fourier transform of the function $B(t)$ as $B(f)$. Given a
function $B(t)=2A(t)\cos\phi(t)$, where $d\ln A/dt\ll d\phi(t)/dt$
and $|d^2\phi/dt^2|\ll (d\phi/dt)^2$, the SPA provides the
following estimate of the Fourier transformation ${B}(f)$ \cite{gw-book,zhao-2017}:
\begin{equation}
{B}(f)\simeq \frac{A(t_f)}{\sqrt{\dot{F}(t_f)}}
e^{i[\Psi_f(t_f)-\pi/4]}, ~~f\ge 0,
\end{equation}
where $\Psi_f(t)\equiv 2\pi ft-\phi(t)$, $2\pi F(t)\equiv
d\phi/dt$. In this formula, $t_f$ is defined as the time at which
$F(t_f)=f$ and $\Psi_f(t_f)$ is the value of $\Psi_f(t)$ at
$t=t_f$. Employing SPA, the Fourier
transform of the time-series data from the $I$-th GW detector can
be obtained as follows,
\begin{equation}
d_I(f)=\int_0^T d_I(t) e^{2\pi ift} dt.
\end{equation}
Denoting the corresponding one-side noise spectral density by
$S_I(f)$, we define a whitened data set in the frequency domain \cite{algorithm2},
\begin{equation}
\hat{d}_I(f)\equiv S_{I}^{-1/2}(f)d_{I}(f).
\end{equation}

For a detector network, this can be rewritten as \cite{algorithm2},
\begin{equation}\label{response}
{\rm {\bf \hat{d}}}(f)=e^{-i {\Phi}} \hat{\rm{\bf
A}}{\rm {\bf h}}(f),
\end{equation}
where $\Phi$ is the $N_d\times N_d$ diagonal matrix with
$\Phi_{IJ}=2\pi f\delta_{IJ}({{\rm{\bf n}}\cdot{{\rm{\bf
r}}_I}}(f))$, and \begin{equation} \hat{\rm{\bf A}}{\rm {\bf
h}}(f)=\left[\frac{F_1^{+}h_+(f)+F_1^{\times}h_{\times}(f)}{\sqrt{S_1(f)}},
\frac{F_2^{+}h_+(f)+F_2^{\times}h_{\times}(f)}{\sqrt{S_2(f)}},
\cdot\cdot\cdot,\frac{F_{N_d}^{+}h_+(f)+F_{N_d}^{\times}h_{\times}(f)}{\sqrt{S_{N_d}(f)}}
\right]^{T}.
\end{equation}
Note that, $F_I^{+}$, $F_I^{\times}$, $\Phi_{ij}$ are all functions with respective to frequency in general, due to the diurnal motion of the Earth. With SPA, these functions are simply given by
\begin{equation}
F_{I}^{+}(f)=F_{I}^{+}(t=t_f), ~~~~F_{I}^{\times}(f)=F_{I}^{\times}(t=t_f), ~~~\Phi_{ij}(f)=\Phi_{ij}(t=t_f),
\end{equation}
where $t_f=t_c-({5}/{256})\mathcal{M}_c^{-5/3}(\pi f)^{-8/3}$ \cite{gw-book}\footnote{Note that, this relationship is untenable in the general modified gravities \cite{zhao-2017}.}, $\mathcal{M}_c$ is the chirp mass of binary system, which will be defined in next paragraph, and $t_c$ is the binary coalescence time. If the effect of the Earth's rotation is ignored, these can be approximately treated as constants for a given GW event. In this paper, we consider both cases to show the effect of the Earth's rotation.

In general, the spinning inspiral-merge-ringdown coalescence waveform template of the compact binaries are needed for the parameter estimation. In order to simplify and speed up the calculation, similar to previous works \cite{sathya2009,zhao-2011,zhao-2017}, in this paper we adopt the restricted post-Newtonian (PN) approximation of the waveform for the non-spinning systems \cite{restricted,restricteds,gw-waveform}, which includes only waveforms in the inspiralling stage. Compared with the case with full waveforms, we do not expect a significant change in our result. For a coalescing binary at a luminosity distance $d_{\rm L}$, with component masses $m_1$ and $m_2$,
total mass $M=m_1+m_2$, symmetric mass ratio $\eta=m_1m_2/M^2$ and ``chirp mass"
$\mathcal{M}_c=M\eta^{3/5}$, the SPA Fourier transform
of a GW waveform is given by \cite{gw-waveform},
\begin{equation}
F_I^{+}h_+(f)+F_I^{\times}h_{\times}(f)=\mathcal{A}_I f^{-7/6}\exp[i(2\pi f
t_c-\pi/4+2\psi(f/2))-\varphi_{I,(2,0)}],
\end{equation}
with the Fourier amplitude $\mathcal{A}_I$ given by
\begin{equation}
\mathcal{A}_I=\frac{1}{d_{\rm
L}}\sqrt{((F_I^+(1+\cos^2\iota))^2+(2F_I^{\times}\cos\iota)^2}\sqrt{5\pi/96}\pi^{-7/6}\mathcal{M}_c^{5/6},
\end{equation}
where $\iota$ is the inclination angle between the binary's orbital angular momentum and the line of sight. We use the 3.5 PN approximation for the phase \cite{restricteds,gw-waveform}, where the functions $\psi$ and $\varphi_{I,(2,0)}$ are given by,
\begin{equation}\label{eqss}
\psi(f)=-\psi_c+\frac{3}{256\eta}\sum_{i=0}^{7}\psi_i(2\pi
Mf)^{i/3},
\end{equation}
\begin{equation}
\varphi_{I,(2,0)}=\tan^{-1}\left(-\frac{2\cos\iota
F_I^{\times}}{(1+\cos^2\iota)F_I^{+}}\right).
\end{equation}
The parameters $\psi_i$ can be found in \cite{gw-waveform}. The upper cutoff
frequency $f_{\rm up}$ is calculated from the last stable orbit, which marks the
end of the inspiral regime and the onset of the final merge. We
will assume that this occurs when the radiation frequency reaches
$f_{\rm up}=2f_{\rm LSO}$, with $f_{\rm LSO}=1/(6^{3/2}2\pi
M)$ the orbital frequency at the last stable orbit. Note that, for GW sources at redshift $z$, the observed mass $m$ is related to the physical (intrinsic) mass $m_{\rm phys}$ by the relation $m=(1+z)m_{\rm phys}$. Throughout this paper, all the masses refer to the observed quantity unless explicitly specified.

%************************************************************************   table 1  ******      table 1  *****************************************************************************************
\begin{table}
\caption{The coordinates of the interferometers used in this
study. Orientation is the smallest angle made by any of the arms
and the local north direction \cite{vilta}.}
\begin{center}
\label{table1}
\begin{tabular}{|c| c| c| c| c|}
    \hline
     & ~~~~$\varphi$~~~~ & ~~~~$\lambda$~~~~ & ~~~~$\gamma$~~~~ &~~~~$\zeta$~~~~ \\
         \hline
    Einstein Telescope in Europe & $43.54^{\circ}$ & $10.42^{\circ}$ & $19.48^{\circ}$ & $60^{\circ}$\\
             \hline
    Cosmic Explorer in the US & $30.54^{\circ}$& $-90.53^{\circ}$  & $162.15^{\circ}$ & $90^{\circ}$\\
             \hline
    Assumed detector in Australia & -31.51$^{\circ}$& $115.74^{\circ}$  & $0^{\circ}$ & --- \\
                 \hline
    Assumed detector in China & 38.39$^{\circ}$& $104.28^{\circ}$  & $89.95^{\circ}$  & --- \\
  \hline
\end{tabular}
\end{center}
\end{table}
%************************************************************************   table 1  ******      table 1 *****************************************************************************************

%%%%%%%%%%%%%%%%%%%%%%%%%%%%%%%%%%%%%%%%%  figure 1, figure 1, figure1%%%%%%%%%%%%%%%%%%%%%%%%%%%%%%%%%%%%%%%%%%%%%%%%%%
\begin{figure}
\begin{center}
\centerline{\includegraphics[width=15cm,height=10cm]{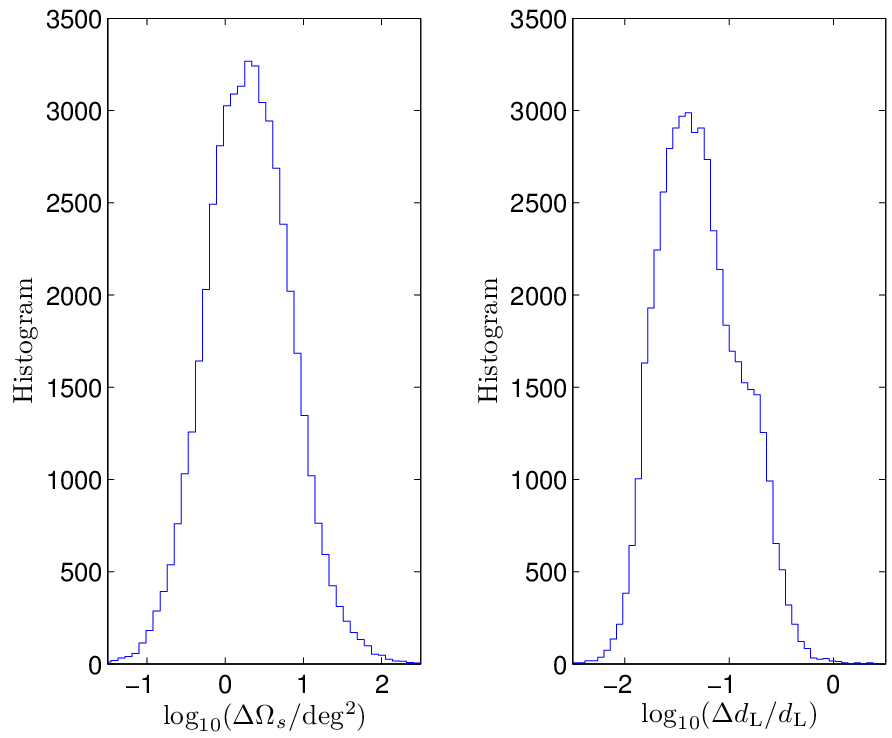}}
\end{center}\caption{The histograms of $\Delta \Omega_s$ (left panel) and $\Delta d_{\rm L}/d_{\rm L}$ (right panel) for $50,000$ BBH samples, which are distributed in the redshift range $z\in[0,3]$ (see the main text for the details). }\label{fig-compare}
\end{figure}
%%%%%%%%%%%%%%%%%%%%%%%%%%%%%%%%%%%%%%%%  figure 1, figure 1, figure1%%%%%%%%%%%%%%%%%%%%%%%%%%%%%%%%%%%%%%%%%%%%%%%%%%

\section{Fisher Information Matrix and Analysis Method \label{section3}}
\label{sec:Fisher}
%{[\r add matched filtering reference]}
By maximizing the correlation between a template waveform that
depends on a set of parameters and a measured signal, the matched filtering provides a natural way to
estimate the parameters of the signal and their errors. With a
given detector noise $S_I(f)$, we employ the Fisher matrix
approach in this paper. In the case of a network including $N_d$ independent GW detectors, the Fisher matrix is given by
\cite{gwfisher},
\begin{equation}\label{fisher}
\Gamma_{ij}=\langle \partial_i {\rm {\bf \hat{d}}}|\partial_j {\rm {\bf \hat{d}}}\rangle,
\end{equation}
where $\partial_i {\rm {\bf \hat{d}}}\equiv\partial {\rm {\bf \hat{d}}}(f)/\partial p_i$, and
$p_i$ denote the free parameters to be estimated. For any given binary system, the response of GW detector in Eq.(\ref{response})
depends on nine system parameters
$({\mathcal{M}}_c,\eta,t_c,\psi_c,\iota,\theta_s,\phi_s,\psi_s,d_{\rm L})$,
where $\psi_c$ is defined in Eq. (\ref{eqss}), and the other parameters are all defined previously.
The angular brackets denote the scalar product, which, for any
two functions $a(t)$ and $b(t)$ is defined as
\begin{equation}
\langle a,b\rangle=2\int_{f_{\rm low}}^{f_{\rm up}}
\left\{\tilde{a}(f)
\tilde{b}^*(f)+\tilde{a}^*(f)\tilde{b}(f)\right\}df.
\end{equation}
Under the assumption of stationary Gaussian detector noise, the optimal squared signal-to-noise ratio (SNR) is given by
\begin{equation}
\rho^2=\langle  {\rm {\bf \hat{d}}}| {\rm {\bf \hat{d}}}\rangle.
\end{equation}

{\r{The Fisher matrix is commonly used in many fields \cite{data-analysis-in-cosmology} to estimate errors in the measured parameters by the expression $\langle \delta p_j \delta p_k\rangle =(\Gamma^{-1})_{jk}$. Once the total
Fisher matrix $\Gamma_{ij}$ is calculated, an estimate of the root mean square (RMS) error,
$\Delta p_{i}$, in measuring the parameter $p_i$ can then be
calculated,  \bea\label{delta-definition} \Delta
p_i=(\Gamma^{-1})_{ii}^{1/2}. \ena The correlation coefficient for any two parameters $p_i$ and $p_j$ is quantified by the ratio $r_{ij}\equiv (\Gamma^{-1})_{ij}/((\Gamma^{-1})_{ii}(\Gamma^{-1})_{jj})^{1/2}$, and the result of $r_{ij}=0$ indicates the no correlation between them. Note that, if any two parameters are completely correlated, i.e. they are degenerate in data analysis,  the Fisher matrix is not invertible. The use of the Fisher matrix to estimate parameter uncertainties is based on the following theorem. The Cramer-Rao bound states that for an unbiased estimator (the
ensemble average of which is the true value), the Fisher matrix
sets a method-independent lower bound for the covariance matrix of
estimated parameters when considering statistical errors \cite{rao}.

In the calculation of Fisher matrix, the quantity $\partial {\rm {\bf \hat{d}}}(f)/\partial p_i$ usually has no analytical expression, and need to be numerically calculated. We adopt the approximation $\partial {\rm {\bf \hat{d}}}(f)/\partial p_i \simeq ({\rm {\bf \hat{d}}}(f;p_i+\delta p_i)-{\rm {\bf \hat{d}}}(f;p_i))/\delta p_i$ and numerically calculate the elements of the Fisher matrix for each GW event. To guarantee the reliability of our calculation, for each GW detector, we vary the values of $\delta p_i$ until this approximation becomes stable, i.e. increasing or decreasing each value of $\delta p_i$ by a factor of $10$, the Fisher matrix has no significant change. Note that the Fisher matrix yields only the lower limit of the covariance matrix.  In reality, a parameter estimation method would have encountered issues such as the well-known multi-modality problem \cite{150914-localization,gw170817} and may or may not reach the limit prescribed by the Fisher matrix.}}

For each GW event observed by the network, we calculate the 9-parameter Fisher matrix, and marginalize
it to the one with two position parameters $(\theta_s,\phi_s)$, which are the colatitude and longitude in a polar coordinate system respectively. The
covariance matrix is given by the inverse of Fisher matrix,
and the error in solid angle (measured in steradians) is given by,
\begin{equation}
\Delta\Omega_s=2\pi|\sin\theta_s|\sqrt{\langle\Delta\theta_s^2\rangle\langle\Delta\phi_s^2\rangle-\langle\Delta\theta_s\Delta\phi_s\rangle^2},
\end{equation}
where $\Delta\theta_s$ and $\Delta\phi_s$ are the deviations of $\theta_s$ and $\phi_s$ from their true values, and the quantities $\langle\Delta\theta_s^2\rangle$, $\langle\Delta\phi_s^2\rangle$ and $\langle\Delta\theta_s\Delta\phi_s\rangle$ are given in the 9-parameter covariance matrix. The uncertainty of $\ln d_{\rm L}$ is straightly derived by employing the formula in Eq. (\ref{delta-definition}). %{\tc{We should stress that, in the realistic analysis, the distributions of model parameters are usually much more complicated. The concrete analysis methods have the bi-modalities problem in usual (for instance, the observed GW150914 event \cite{150914-localization}), and cannot arrive at the limit derived from Fisher matrix analysis. }}

Throughout this paper, we simulate numerically binary coalescence signals with random binary orientations and sky directions for our investigations.
We also generate the random samples at every redshift with a 0.1 spacing within a specified range, and the luminosity distance is calculated with a standard $\Lambda$CDM cosmology \cite{planck} assumed. The sky direction, binary inclination and polarization angle of these binaries are randomly chosen in the angular ranges of $\cos\theta_s\in[-1,1]$, $\phi_s\in[0,360^{\circ}]$, $\psi_s\in[0,360^{\circ}]$ and $\cos\iota\in[-1,1]$.  Without loss of generality, the merger time of these samples are chosen to be $t_c=0$.  The detection criteria is chosen to have the optimal network SNR $\ge 8$.

{
To demonstrate the effectiveness of our Fisher matrix technique, we compare our method to the Markov Chain Monte Carlo (MCMC) analysis used in \cite{vilta}, which analyzed the localization of GW sources with the nested sampling flavor of lalinference \cite{algorithm4} for 3G detectors with $f\ge 10$ Hz.  Similar to \cite{vilta}, we simulate $50,000$ random BBH samples, and assume that the total physical mass of each BBH is uniformly distributed in the range of $[12,40]$ \mSun For the BBHs with larger masses $M_{\rm phys}\gtrsim 40$ \mSun, the high-frequency cutoff $f_{\rm LSO}$ of the restricted PN approximation of the waveform is smaller than 50 Hz and the ring-down signals become important, so the PN approximation becomes unsuitable for 3G detectors. We therefore do not consider these events.\footnote{} with a minimum mass ratio of $1/3$. The redshifts are uniformly distributed in comoving volume, assuming a standard $\Lambda$CDM cosmology \cite{planck}, in the range $z\in[0,3]$. We calculate the 2-detector network SNR values of these binaries for a network of an ET-D in Europe and a CE in the US (Fig.~\ref{f0}), and select the  sources with SNRs in the range of $[10,600]$ as in \cite{vilta}.  Our results are shown in Fig.~\ref{fig-compare}  for the distribution of the angular resolution $\Delta\Omega_s$ and distance uncertainty $\Delta d_{\rm L}/d_{\rm L}$. These results are comparable to that in Fig.~3 and Fig.~4 in \cite{vilta} from the MCMC analysis.  While the Fisher matrix yields a lower limit for the covariance matrix for parameter estimation,  which cannot always be achieved, it represents a simple, method-independent,  and reasonable estimate for the detection and localization capability for future experiments and is especially helpful for comparing results of parameter estimation among different experiments.
}

%The samples a

\section{Effect of low-frequency sensitivity\label{section4}}
\label{sec:One3G}

%%%%%%%%%%%%%%%%%%%%%%%%%%%%%%%%%%%%%%%%%  figure 1, figure 1, figure1%%%%%%%%%%%%%%%%%%%%%%%%%%%%%%%%%%%%%%%%%%%%%%%%%%
\begin{figure}
\begin{center}
\centerline{\includegraphics[width=15cm]{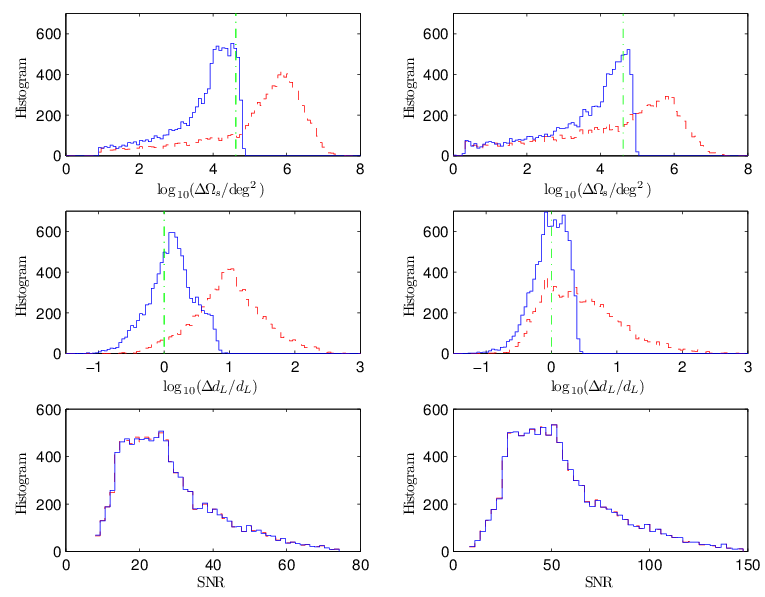}}
\end{center}\caption{{Distribution of {\b the ET-B} angular resolution $\Delta\Omega_s$ (upper), distance determination accuracy $\Delta d_{\rm L}/d_{\rm L}$ (middle) and optimal SNR (lower) with (blue solid lines) and without (red dashed lines) considering the Earth's rotation for the (1.4+1.4) \mSun\, BNS systems (left panels) and for the (1.4+10) \mSun\, NSBH systems (right panels) {at a distance of 1 Gpc}.}}\label{fa1}
\end{figure}
%%%%%%%%%%%%%%%%%%%%%%%%%%%%%%%%%%%%%%%%  figure 1, figure 1, figure1%%%%%%%%%%%%%%%%%%%%%%%%%%%%%%%%%%%%%%%%%%%%%%%%%%

%%%%%%%%%%%%%%%%%%%%%%%%%%%%%%%%%%%%%%%%%  figure 1, figure 1, figure1%%%%%%%%%%%%%%%%%%%%%%%%%%%%%%%%%%%%%%%%%%%%%%%%%%
\begin{figure}
\begin{center}
\centerline{\includegraphics[width=15cm]{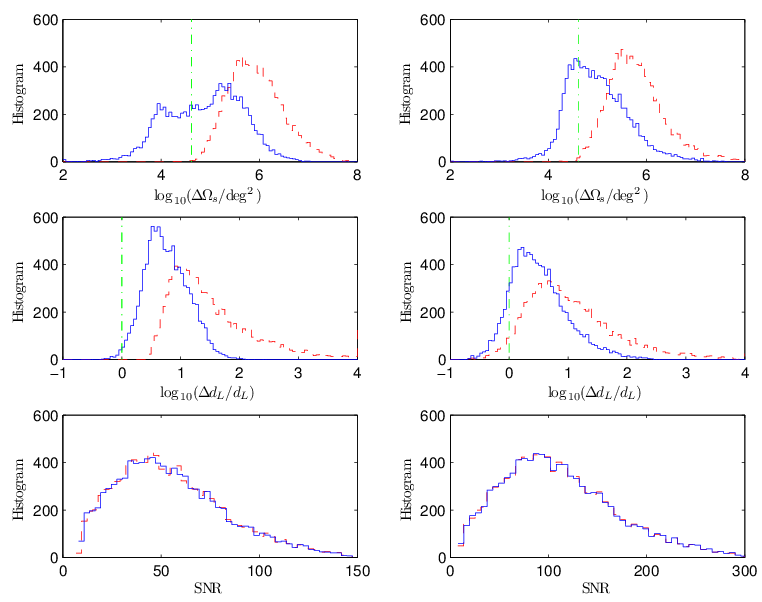}}
\end{center}\caption{Same as Fig.~\ref{fa1}, but {\b for CE}.}\label{fa11}
\end{figure}
%%%%%%%%%%%%%%%%%%%%%%%%%%%%%%%%%%%%%%%%  figure 1, figure 1, figure1%%%%%%%%%%%%%%%%%%%%%%%%%%%%%%%%%%%%%%%%%%%%%%%%%%

%%%%%%%%%%%%%%%%%%%%%%%%%%%%%%%%%%%%%%%%%  figure 1, figure 1, figure1%%%%%%%%%%%%%%%%%%%%%%%%%%%%%%%%%%%%%%%%%%%%%%%%%%
\begin{figure}
\begin{center}
\centerline{\includegraphics[width=15cm]{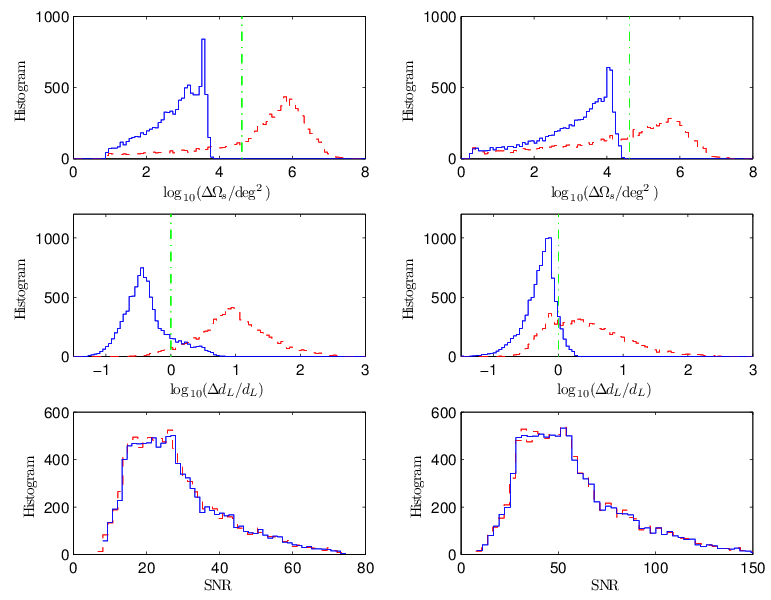}}
\end{center}\caption{Same as  Fig.~\ref{fa1}, but {\b for ET-D}.}\label{fb1}
\end{figure}
%%%%%%%%%%%%%%%%%%%%%%%%%%%%%%%%%%%%%%%%  figure 1, figure 1, figure1%%%%%%%%%%%%%%%%%%%%%%%%%%%%%%%%%%%%%%%%%%%%%%%%%%

% ideal detector's single detector at 1 Gpc
%\begin{figure}
%\begin{center}
%\centerline{\includegraphics[width=15cm]{figs2/single/hist/ideal/NSNS_NSBH_3_SNR8_2.png}}
%\end{center}\caption{Same with Fig.~\ref{fa1}, but for the ideal detector described in Fig~\ref{f0}.}\label{fc1}
%\end{figure}

%%%%%%%%%%%%%%%%%%%%%%%%%%%%%%%%%%%%%%%%%  figure 1, figure 1, figure1%%%%%%%%%%%%%%%%%%%%%%%%%%%%%%%%%%%%%%%%%%%%%%%%%%
\begin{figure}
\begin{center}
\centerline{\includegraphics[width=15cm]{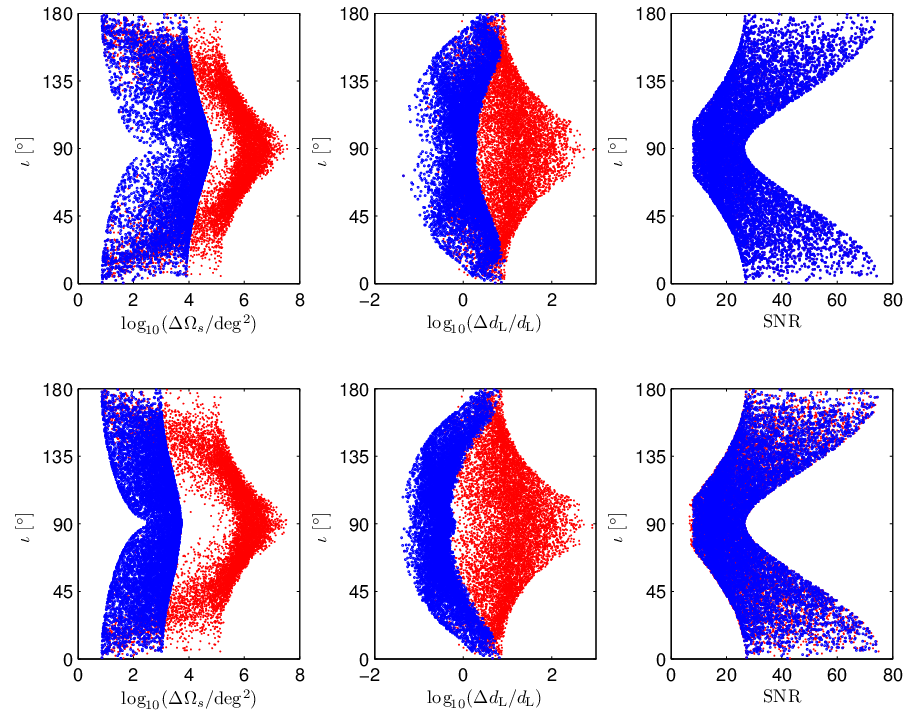}}
\end{center}\caption{ The distributions of $\Delta \Omega_s$ (left), $\Delta d_{\rm L}/d_{\rm L}$ (middle) and SNR (right) with respective to the inclination angle $\iota$ for $10^4$ BNS samples at the distance $d_{\rm L}=1$ Gpc. The upper panels show the results of ET detector with {\b ET-B noise}, and the lower ones show those of ET detector with {\b ET-D noise}. In each panel, the red dots indicate the distribution of samples in the case without considering the Earth's rotation, and the blue circles indicate that in the case with considering the Earth's rotation.}\label{fz1}
\end{figure}
%%%%%%%%%%%%%%%%%%%%%%%%%%%%%%%%%%%%%%%%  figure 1, figure 1, figure1%%%%%%%%%%%%%%%%%%%%%%%%%%%%%%%%%%%%%%%%%%%%%%%%%%

%%%%%%%%%%%%%%%%%%%%%%%%%%%%%%%%%%%%%%%%%  figure 1, figure 1, figure1%%%%%%%%%%%%%%%%%%%%%%%%%%%%%%%%%%%%%%%%%%%%%%%%%%
\begin{figure}
\begin{center}
\centerline{\includegraphics[width=15cm]{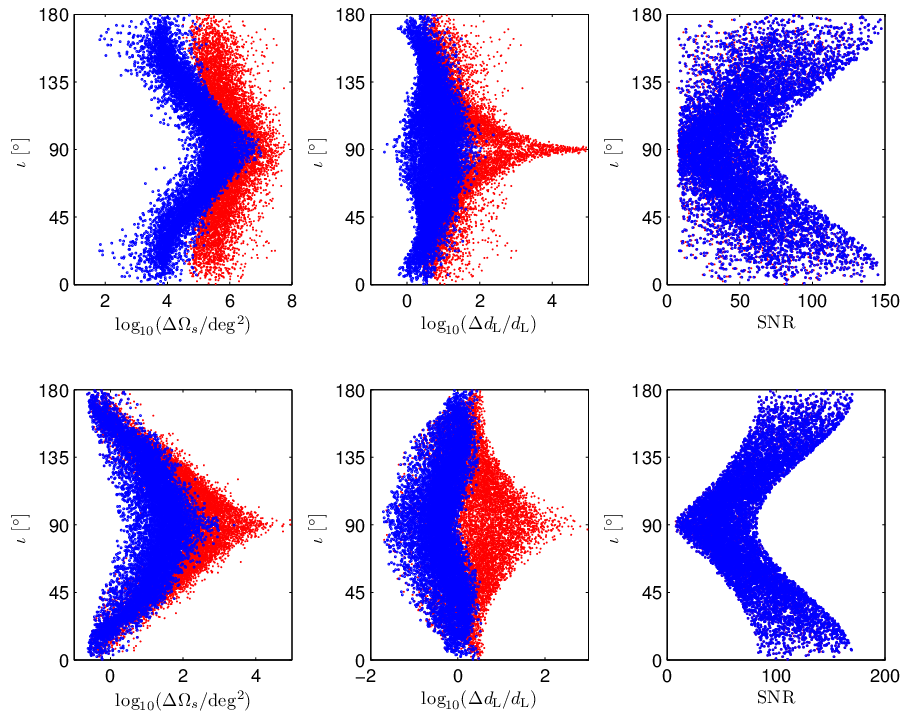}}
\end{center}\caption{Same as Fig.~\ref{fz1}, but the upper panels show the results of {\b CE}, and the lower panels show those of {\b the 2CE detector network}.}\label{fz2}
\end{figure}
%%%%%%%%%%%%%%%%%%%%%%%%%%%%%%%%%%%%%%%%  figure 1, figure 1, figure1%%%%%%%%%%%%%%%%%%%%%%%%%%%%%%%%%%%%%%%%%%%%%%%%%%

% discuss the principle first

Consideration of the angular resolution of a 3G detector is different from a 2G detector in two possible aspects besides the increased SNRs due to better sensitivity.  (1) An ET-like detector contains three co-located independent V-shaped Michelson interferometers with an opening angle of $60^{\circ}$ and rotated relative to each other by 120$^{\circ}$. {The three V-shaped interferometers are equivalent, in terms of antenna beam-pattern and sensitivity, to two co-located L-shaped interferometers whose arms are three-quarters in length and rotated relative to each other by an angle of  $45^{\circ}$. } It has essentially no blind spots on sky \cite{mock}. Therefore, the directional sensitivity is greatly improved \cite{ref:3Gdesign} than one L-shaped 2G detector.  In addition,  GW polarizations and polarization angle  \cite{ref:3Gdesign} can  be determined by one 3G detector. (2) Extended lower frequency sensitivity is expected for the 3G detectors \cite{3G-lowfreq}.  Therefore, the observed GW signal duration is possibly a significant fraction of the Earth's rotation period.  The angular resolution of the detector can therefore be improved by the non-negligible GW energy-flux weighted geometrical area formed by the trajectory of the detector due to the Earth self rotation within the signal duration. This is similar to the case for the angular resolution for the space-based detector LISA as discussed in Wen \& Chen (2010) \cite{algorithm2}, (see also in \cite{LISA1,LISA2}).

For binary coalescence, the duration of the signal $t_*$  in a detector band is a strong function of the detector's low-frequency cutoff $f_{\rm low}$ \cite{gw-book},
\begin{equation}\label{t_star}
t_*=0.86~{\rm day} \left(\frac{1.21 ~{\rm M}_{\odot}}{\mathcal{M}_c}\right)^{5/3} \left(\frac{2~{\rm Hz}}{f_{\rm low}}\right)^{8/3},
\end{equation}
where $\mathcal{M}_c$ is the chirp mass of the system. For the BNS with $m_1=m_2=1.4$ \mSun, we have $t_*=0.28$ hours for $f_{\rm low}=10~{\rm Hz}$, $t_*=0.29$ days for $f_{\rm low}=3{\rm~Hz}$, and $t_*=5.44$ days for $f_{\rm low}=1{\rm~Hz}$. This evaluation shows that if the sensitivity for GWs at $f\lesssim 10{\rm~Hz}$ is non-negligible, the Earth's rotation will play a crucial role for the localization of GW sources. For sources with higher masses (e.g., BBH, NSBH), the effect of Earth's rotation becomes important at lower $f_{\rm low}$ than the BNS systems. However, the duration of GW signals in the detector frequency band is too short for the low-$f$ effect to be significant. For instance, for a BBH system with $m_1=m_2=30$ \mSun, which is similar to that of GW150914, we have $t_*=0.79$ hours for $f_{\rm low}=1$ Hz, therefore the effect of the Earth's rotation is expected to be small. For this reason, in this article, we only consider the compact binary systems of BNSs and NSBHs with small black hole masses.

\subsection{Effect of the Earth's rotation}

We first demonstrate the importance of considering the {time-dependent antenna beam-pattern function due to the Earth's rotation} when calculating the angular resolution and distance uncertainties of BNSs and NSBHs for 3G detectors with the proposed design of ET-B, CE and ET-D.   {To illustrate our point, we compare the angular resolution between the following two cases when calculating the Fisher matrix $\Gamma_{ij}$. (1) Constant antenna beam-pattern functions and time delay between detectors, that is fixing the $F_I^{+}$, $F_I^{\times}$ and $\Phi$ values in Eq. (\ref{response}) at $t=0$ (Fig.~\ref{fa1}-Fig.~\ref{fb1}, red dashed lines), and (2) $t$-dependent (or $f$-dependent) $F_I^{+}$, $F_I^{\times}$ and $\Phi$ in Eq. (\ref{response}) (Fig.~\ref{fa1}-Fig.~\ref{fb1}, blue solid lines).
%\[
%\left \{
%\begin{tabular}{ccc}
%{\bf Case A:}  &  &  \, \, \, constant $F_I^{+}$, $F_I^{\times}$ and $\Phi$ in Eq. (\ref{response}) by fixing them at $t=0$,\\
%{\bf Case B:}  &  &t-dependent (or f-dependent) $F_I^{+}$, $F_I^{\times}$ and $\Phi$ in Eq. (\ref{response}).
%\end{tabular}
%\right.
%\]
%{\bf Case A:} In the calculation of Fisher matrix $\Gamma_{ij}$, we ignore the time dependence (equivalently, frequency dependence) of the functions $F_I^{+}$, $F_I^{\times}$ and $\Phi$ in Eq. (\ref{response}) by fixing them at $t=0$.
%{\bf Case B:} We take into account the rotation of the Earth, and the functions $F_I^{+}$, $F_I^{\times}$ and $\Phi$ are time dependent, or equivalently, frequency dependent in Fourier domain.
The comparisons are conducted for the detected BNS and NSBH sources from a set of $10^4$ randomly sampled (Sec.~\ref{sec:Fisher}) BNS and NSBH sources respectively at a luminosity distance of 1 Gpc.  %The results are shown in Figs.~\ref{fa1}-\ref{fb1}, for ET-B, CE and ET-D, respectively.

We find that for the BNS systems (Fig.~\ref{fa1}, left panels), without considering the time-dependence of the antenna beam-pattern functions (Fig.~\ref{fa1}, red dashed lines), ET-B alone cannot localize most of the GW sources at this distance, even at high SNRs.   %Fig.~\ref{fa1} shows that $\Delta\Omega_s\> 4\times 10^4{\rm deg}^2$ and $\Delta d_{\rm L}/d_{\rm L}>$ a few for $>50$\% of BNS sources.
However, if taking into account the time-dependence (Fig.~\ref{fa1}, blue solid lines), while the SNR values are largely unchanged (Fig.~\ref{fa1}, bottom panels), around $20\%$ BNS sources can be localized with $\Delta\Omega_s<10^3 ~{\rm deg}^2$, and $14\%$ sources have a distance uncertainty $\Delta d_{\rm L}/d_{\rm L}<50$\%.  Similar result can be found for the NSBH systems but with less significant effect due to the overall shorter duration of these sources (Fig.~\ref{fa1}, right panels).

The effect of considering time-dependent antenna beam-pattern functions for a CE can be found in Fig.~\ref{fa11} for the same sample of sources.  The SNR values (lower panels) are significantly larger for CE than those for the ET-B, due to its much better sensitivity around 100 Hz.  The effect of the Earth's rotation on CE is still noticeable, but much less pronounced (upper and middle panels) than ET-B,  as CE has a poorer sensitivity in the sub-$5$ Hz low-frequency range than ET-B.  Both the angular resolution and distance accuracy of the BNS and NSBH detected by the CE are overall much worse than ET-B despite the much larger SNRs. This is understandable as CE is designed to be very similar to a 2G detector,  but with a much improved sensitivity. It therefore has similarly poor single-detector directional sensitivity as a 2G detector.  In comparison,  ET-B has the advantages of being equivalent to two L-shaped detectors plus better low frequency sensitivity.  In other words, the time-dependence of the antenna beam-pattern function and the design of ET help better determine the wave polarizations and merger time which in turn help the source localization.
}

The influence of the Earth's rotation is much more prominent in ET-D (Fig.~\ref{fb1}) than in ET-B, as expected from its much lower noise level at low frequencies of $f\in(2,20)$ Hz.  The angular resolutions of $\Delta\Omega_s$ in ET-D are one order of magnitude smaller than in ET-B, and the distance determination accuracy $\Delta d_{\rm L}/d_{\rm L}$ is nearly three times better.  Nearly 50\% BNS systems can be localized within 1000 deg$^2$ at $d_{\rm L}=1$ Gpc with one single ET-D. This, when scaled with distance, is comparable to the localization accuracy of their first detected GW source GW150914 by the two Advanced LIGO detectors during the first science run. Around 20\% of BNSs can possibly be localized within 100 deg$^2$ by an ET-D while the distance accuracy is well within 30\%.  For NSBH systems, the angular resolution and distance determination are slightly worse than the BNS systems.

{The effect of the Earth's rotation on the dependence of the angular resolution, the distance uncertainty and SNR with the binary inclination angle $\iota$ is also shown for the same $10^{4}$ BNS samples at 1 Gpc for an ET-B  (Fig.~\ref{fz1}, upper panels), an ET-D (Fig.~\ref{fz1}, lower panels), a CE detector (Fig.~\ref{fz2}, upper panels), and for two CE detectors in Europe and the US respectively (Fig.~\ref{fz2}, lower panels). The angular resolution of face-on ($\iota \sim 0^{\circ}$ or $180^{\circ}$) binaries is overall better than those edge-on systems ($\iota \sim 90^{\circ}$) due to larger SNR values. This is desirable for EM observations of BNS systems associated with on-axis gamma-ray bursts (GRBs).  On the other hand,  the time-dependent antenna beam-pattern functions can help better localize those edge-on systems whose GRB counterparts are less likely to be observed at high SNRs as the GRB jet opening angles are expected to be small. Such improvement is much more pronounced in the ETs than in CE due to ET's better low frequency sensitivity.}

{\r{ From Figs.~\ref{fa1} -\ref{fb1}, we find that there is a significant fraction of samples, where both the sky position and the luminosity distance cannot be resolved, even if we consider the effect of the Earth's rotation. Note that these results are derived from the GW observation alone. If one can identify the EM counterparts of GW events by the coincidence of their arrival times, the sky position $\theta_s$, $\phi_s$ (and $\iota$, $\psi$ for the face-on sources) can be determined in advance, the errors of luminosity distance $d_{\rm L}$ can then be significantly reduced \cite{zhao-2011} and the bi-modal distribution of parameters in realistic analysis could be solved \cite{gw170817}. }}

\subsection{Dependence of angular resolution and distance accuracy on $f_{\rm low}$}
{The effect of the Earth's rotation on the angular resolution of a 3G detector is similar to that of the space-based detector LISA \cite{algorithm2,LISA1,LISA2}.  For LISA, the angular resolution follows roughly a broken power of the observation time with turn-around time near the period of the Earth around the Sun.  This is because the angular resolution is inversely proportional to the projected geometrical area weighted by GW flux in the direction normal to the wave direction formed by the trajectory of the detector. Once the signal is much longer than the Earth's rotation period around the Sun, the project area is approximately a constant therefore the angular resolution improves mainly due to accumulated SNRs. For a 3G detector, as the rotation of the Earth is important, we expect to observe similar approximate broken power-law dependence of the angular resolution to the observing duration, or in this case, the low-frequency cutoff. The turn-around frequency is expected to be around the frequency when the signal duration is comparable to the Earth self-rotation period.}

%%%%%%%%%%%%%%%%%%%%%%%%%%%%%%%%%%%%%%%%%  figure 1, figure 1, figure1%%%%%%%%%%%%%%%%%%%%%%%%%%%%%%%%%%%%%%%%%%%%%%%%%%
\begin{figure}
\begin{center}
\centerline{\includegraphics[width=17cm]{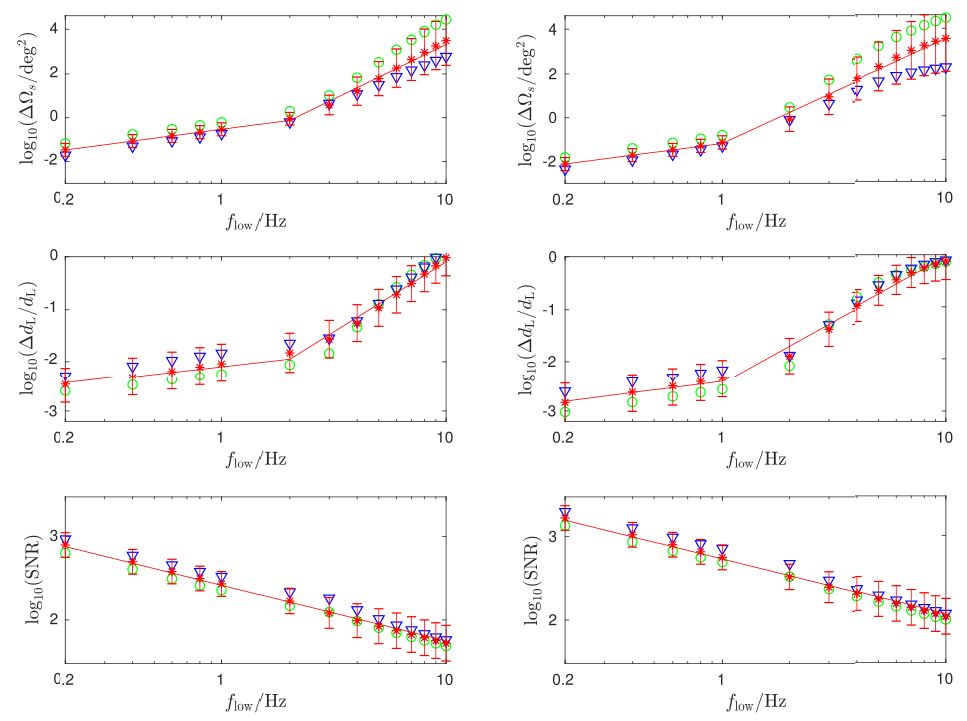}}
\end{center}\caption{Angular resolution $\Delta \Omega_s$, distance accuracy $\Delta d_{\rm L}/d_{\rm L}$ and SNR as a function of the low-frequency cutoff $f_{\rm low}$ for {\b the ideal experiment}.  The left panels show the results of BNS sources, and the right panels show the results of NSBH sources. In each panel, the green circles are for a source at $\theta_s=110.38^{\circ}$, $\phi_s=291.48^{\circ}$, $\psi_s=188.36^{\circ}$, $\iota=75.70^{\circ}$, and the blue triangles are for a GW source at $\theta_s=30.37^{\circ}$, $\phi_s=118.05^{\circ}$, $\psi_s=21.45^{\circ}$, $\iota=46.11^{\circ}$, randomly chosen from $10^4$ samples. The red asterisks and the error bars show the median values and the corresponding standard deviations derived from $10^4$ random samples. Solid red lines are the best-fit broken power-law forms.}\label{fs2}
\end{figure}
%%%%%%%%%%%%%%%%%%%%%%%%%%%%%%%%%%%%%%%%  figure 1, figure 1, figure1%%%%%%%%%%%%%%%%%%%%%%%%%%%%%%%%%%%%%%%%%%%%%%%%%%

%%%%%%%%%%%%%%%%%%%%%%%%%%%%%%%%%%%%%%%%%  figure 1, figure 1, figure1%%%%%%%%%%%%%%%%%%%%%%%%%%%%%%%%%%%%%%%%%%%%%%%%%%
\begin{figure}
\begin{center}
\centerline{\includegraphics[width=17cm]{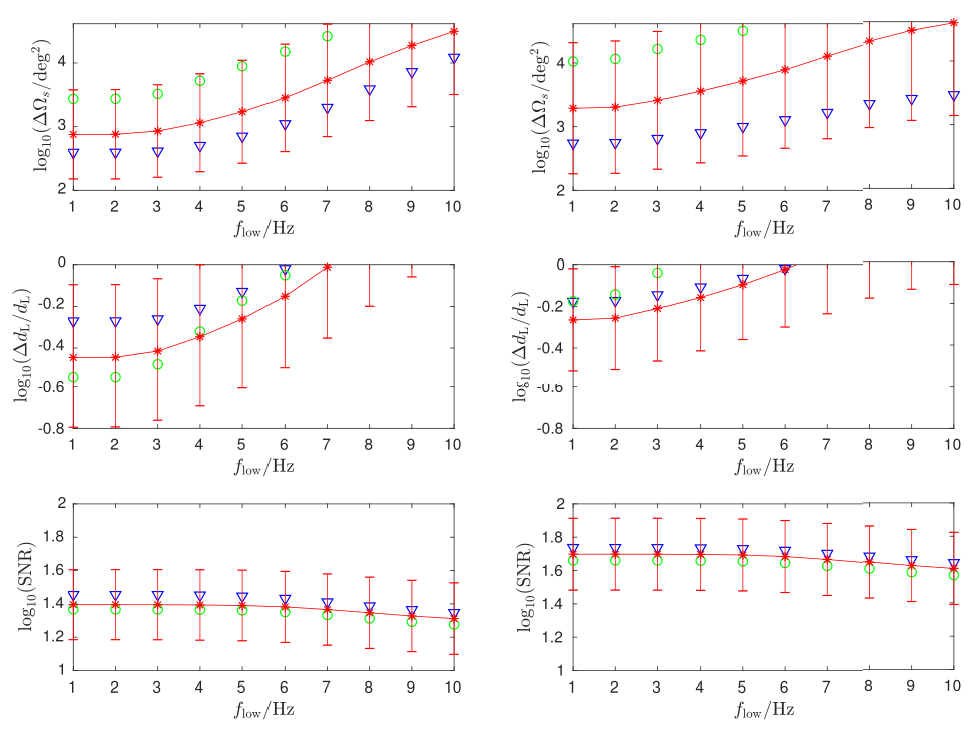}}
\end{center}\caption{The same as Fig.~\ref{fs2} but with {\b the ET-D} sensitivity for different low-frequency cutoff $f_{\rm low}$ between 1 and 10 Hz.
%Different from Fig.\ref{fs2}, the solid red lines here are not the fitting relations.
}\label{fs1}
\end{figure}
%%%%%%%%%%%%%%%%%%%%%%%%%%%%%%%%%%%%%%%%  figure 1, figure 1, figure1%%%%%%%%%%%%%%%%%%%%%%%%%%%%%%%%%%%%%%%%%%%%%%%%%%

%\subsection{Dependence on low-frequency cut localization of GW sources by an individual detector}

{As a proof of principle, we first demonstrate the effect of $f_{\rm low}$ on the angular resolution and distance determination using an ideal detector (Fig.~\ref{fs2}) similar to that described in Fig.~\ref{f0} but with flat noise level extended to the entire frequency band.  We show the results for BNS (left panels) and NSBH (right panels) at 1 Gpc at two randomly chosen sky directions (Fig.~\ref{fs2}, green circles and blue triangles) with random binary orientations, as well as the median values of a random sample of $10^4$ sources over different sky directions and binary orientations (Fig.~\ref{fs2}, red asterisks).  A pronounced broken power-law relation can be observed for the angular resolution and distance determination independent of the source sky directions.  The turn-around frequencies for the broken power-law are around $2$ Hz, and $1$ Hz for the BNS and NSBH sources respectively, around which the signal duration is comparable to the Earth's rotation period. Specifically, we found for the BNS systems,  }
%\begin{eqnarray}
%\log_{10}(\Delta\Omega_s/{\rm deg^2})&=&
%\begin{cases}
%-0.090+1.40(x-x_0), & f\le2{\rm Hz}, \\
%-0.090+2.62(x-x_0)+6.45(x-x_0)^2-4.87(x-x_0)^3, & f>2{\rm Hz},
%\end{cases}\\
%\log_{10}(\Delta d_{\rm L}/d_{\rm L})&=&
%\begin{cases}
%-1.70+0.63(x-x_0), & f\le2{\rm Hz}, \\
%-1.70+0.68(x-x_0)+4.53(x-x_0)^2-3.04(x-x_0)^3, & f>2{\rm Hz},
%\end{cases}\\
%\log_{10}({\rm SNR})&=&
%\begin{cases}
%2.23-0.67(x-x_0), & f\le2{\rm Hz}, \\
%2.23-0.88(x-x_0)+0.30(x-x_0)^2-0.14(x-x_0)^3, & f>2{\rm Hz},
%\end{cases}
%\end{eqnarray}
\begin{eqnarray}
\log_{10}(\Delta\Omega_s/{\rm deg^2})&=&
\begin{cases}
-0.12+1.36(x-x_0), & f\le2{\rm Hz}, \\
 -0.12+4.96(x-x_0), & f>2{\rm Hz},
\end{cases}\\
\log_{10}(\Delta d_{\rm L}/d_{\rm L})&=&
\begin{cases}
-1.95+0.50(x-x_0), & f\le2{\rm Hz}, \\
-1.95+2.68(x-x_0), & f>2{\rm Hz},
\end{cases}
%\log_{10}({\rm SNR})&=&
%\begin{cases}
%2.22-0.68(x-x_0), & ~~f\le2{\rm Hz}, \\
%2.22-0.74(x-x_0), & ~~f>2{\rm Hz},
%\end{cases}
\end{eqnarray}
and for NSBH,
%\begin{eqnarray}
%\log_{10}(\Delta\Omega_s/{\rm deg^2})&=&
%\begin{cases}
%-1.15+1.40(x-x_0), & f\le1{\rm Hz}, \\
%-1.15+1.71(x-x_0)+7.34(x-x_0)^2-4.83(x-x_0)^3, & f>1{\rm Hz},
%\end{cases}\\
%\log_{10}(\Delta d_{\rm L}/d_{\rm L})&=&
%\begin{cases}
%-2.21+0.68(x-x_0), & f\le1{\rm Hz}, \\
%-2.21+0.21(x-x_0)+0.45(x-x_0)^2-2.85(x-x_0)^3, & f>1{\rm Hz},
%\end{cases}\\
%\log_{10}({\rm SNR})&=&
%\begin{cases}
%2.75-0.67(x-x_0), & f\le1{\rm Hz}, \\
%2.75-0.85(x-x_0)+0.22(x-x_0)^2-0.096(x-x_0)^3, & f>1{\rm Hz},
%\end{cases}
%\end{eqnarray}
\begin{eqnarray}
\log_{10}(\Delta\Omega_s/{\rm deg^2})&=&
\begin{cases}
-1.17+1.35(x-x_0), & f\le1{\rm Hz}, \\
-1.17+4.71(x-x_0), & f>1{\rm Hz},
\end{cases}\\
\log_{10}(\Delta d_{\rm L}/d_{\rm L})&=&
\begin{cases}
-2.42+0.56(x-x_0), & f\le1{\rm Hz}, \\
-2.42+2.39(x-x_0), & f>1{\rm Hz},
\end{cases}
%\log_{10}({\rm SNR})&=&
%\begin{cases}
%2.75-0.69(x-x_0), & ~~f\le1{\rm Hz}, \\
%2.75-0.73(x-x_0), & ~~f>1{\rm Hz},
%\end{cases}
\end{eqnarray}
where $x\equiv \log_{10}(f/{\rm Hz})$ and $x_0$ are the $x$ values at the turning-around point of $f=2~{\rm Hz}$ and $f= 1 ~{\rm Hz}$ for BNS and NSBH respectively.   The median values of SNR (Fig.~\ref{fs2}, red asterisks) for both sources roughly follow the power-law relation of ${\rm SNR}\propto f_{\rm low}^{-2/3}$ (Fig.~\ref{fs2}, red solid line), expected for the binary inspiral signal with a flat noise spectrum density and a constant antenna beam-pattern functions.  Results of individual sources (Fig.~\ref{fs2} and Fig.~\ref{fs1}, green circles and blue triangles) can be observed to deviate slightly from this power-law between 2-10 Hz due to the time-dependence of the antenna beam-pattern functions.

%ET-D
{Similar dependence of the angular resolution and low-frequency cutoff can be found in ET-D (Fig.~\ref{fs1}), where we have used the same samples of BNS and NSBH sources as for the ideal detector for our investigation.  For the BNSs, we can observe a relatively steep change in angular resolution and distance accuracy with the low-frequency cutoff at $f_{\rm low}>2$ Hz, while the values of SNR change very little. Specifically, for $f_{\rm low}\gtrsim 3$ Hz, the angular resolution can be approximated by the power-law relations of  $\Delta\Omega_s\propto f_{\rm low}^{2.2}$ for BNSs and $\Delta\Omega_s\propto f_{\rm low}^{1.8}$ for NSBHs. The power-law index is different from that of the ideal detector, due to the frequency-dependence of the ET-D sensitivity curve. } {Consistently, at $f_{\rm low}\gtrsim 10$ Hz, a single detector will have difficulty in localizing GW sources. Note the SNR values in ET-D remain nearly constant for different $f_{\rm low}$ values.

% implication to detector design and data analysis
Our results show that in terms of source localization of single 3G detectors, it is worthwhile to extend the detector's sensitivity to frequencies below $10$ Hz. For the ET-D type of sensitivity, analyzing data at frequencies lower than 10 Hz could help improve the source localization accuracy even if it does not help increase the SNRs. }

\section{Angular resolution and distance determination as a function of redshift}

%%%%%%%%%%%%%%%%%%%%%%%%%%%%%%%%%%%%%%%%%  figure 1, figure 1, figure1%%%%%%%%%%%%%%%%%%%%%%%%%%%%%%%%%%%%%%%%%%%%%%%%%%
\begin{figure}
\begin{center}
\centerline{\includegraphics[width=15cm]{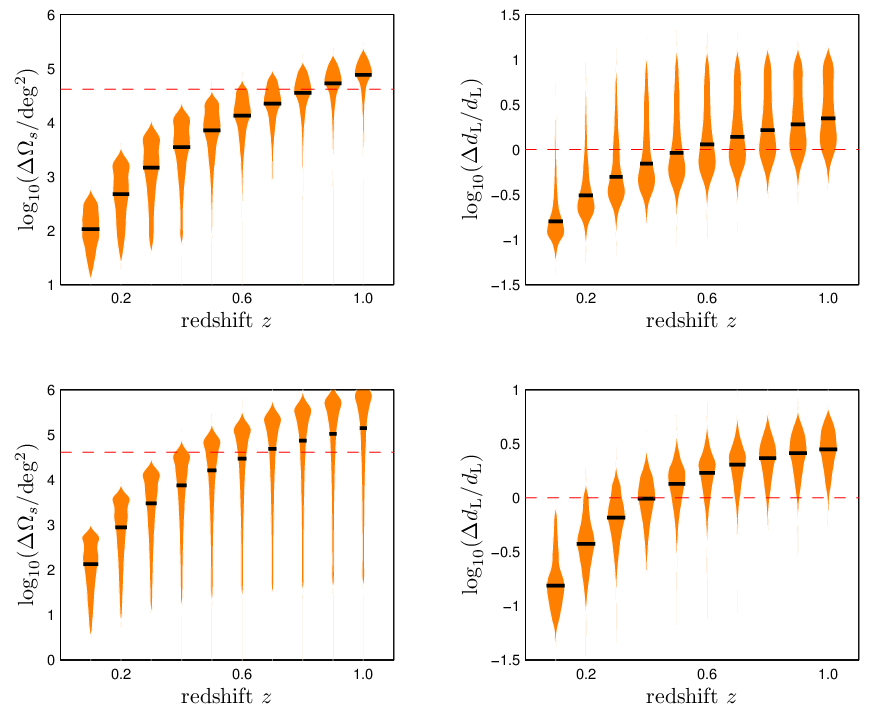}}
\end{center}\caption{Violin plots of the angular resolution $\Delta\Omega_s$ (left panels) and distance uncertainties $\Delta d_{\rm L}/d_{\rm L}$ (right panels) for {\b the ET-D} as a function of redshift $z$, for BNSs (upper panels) and NSBHs (lower panels). The black bars indicate the median values of the corresponding distribution. }\label{ft1}
\end{figure}
%%%%%%%%%%%%%%%%%%%%%%%%%%%%%%%%%%%%%%%%  figure 1, figure 1, figure1%%%%%%%%%%%%%%%%%%%%%%%%%%%%%%%%%%%%%%%%%%%%%%%%%%

%%%%%%%%%%%%%%%%%%%%%%%%%%%%%%%%%%%%%%%%%  figure 1, figure 1, figure1%%%%%%%%%%%%%%%%%%%%%%%%%%%%%%%%%%%%%%%%%%%%%%%%%%
\begin{figure}
\begin{center}
\centerline{\includegraphics[width=15cm,height=10cm]{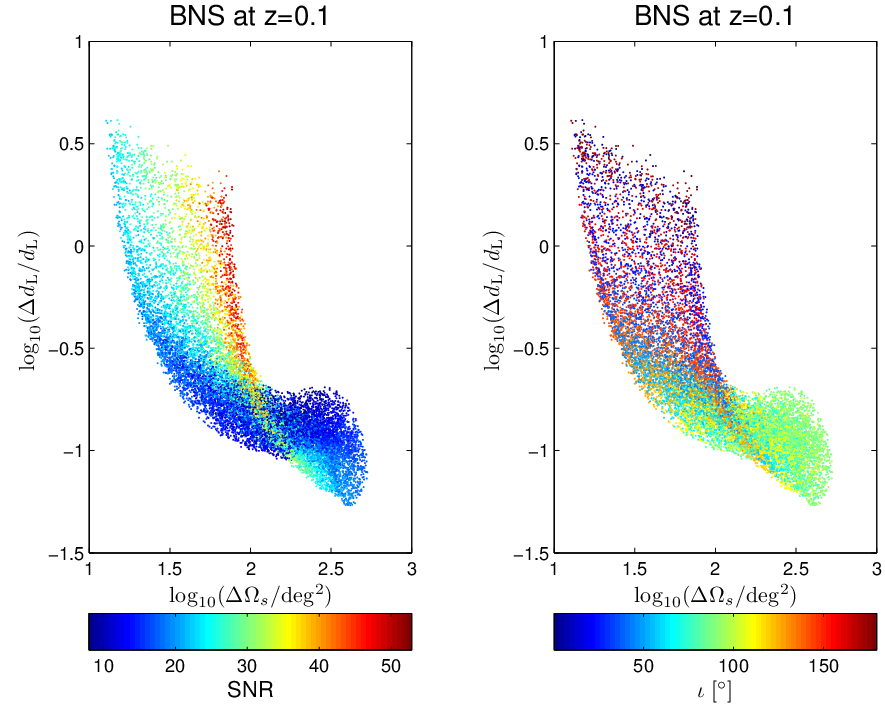}}
\end{center}\caption{The distributions of $\Delta d_{\rm L}/d_{\rm L}$ with respective to $\Delta \Omega_s$ for $15,000$ BNS samples at redshift $z=0.1$. The figure shows the results of ET detector located at Europe with {\b ET-D noise}. In the left panel, the color indicates the SNR value of the sample, and in the right panel it indicates the inclination angle $\iota$.}\label{fz3}
\end{figure}
%%%%%%%%%%%%%%%%%%%%%%%%%%%%%%%%%%%%%%%%  figure 1, figure 1, figure1%%%%%%%%%%%%%%%%%%%%%%%%%%%%%%%%%%%%%%%%%%%%%%%%%%

%%%%%%%%%%%%%%%%%%%%%%%%%%%%%%%%%%%%%%%%%  figure 1, figure 1, figure1%%%%%%%%%%%%%%%%%%%%%%%%%%%%%%%%%%%%%%%%%%%%%%%%%%
\begin{figure}
\begin{center}
\centerline{\includegraphics[width=15cm]{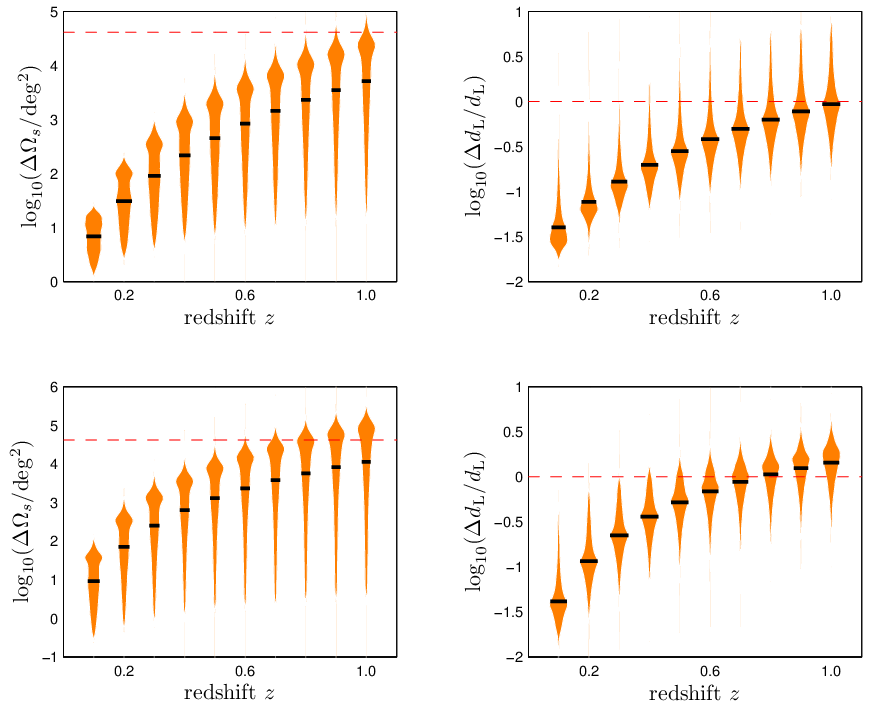}}
\end{center}\caption{Same as Fig.~\ref{ft1}, but ET-D noise is replaced by {\b the ideal noise} described in Fig.~\ref{f0} with low-frequency cutoff $f_{\rm low}=5$ Hz.}\label{ft2}
\end{figure}
%%%%%%%%%%%%%%%%%%%%%%%%%%%%%%%%%%%%%%%%  figure 1, figure 1, figure1%%%%%%%%%%%%%%%%%%%%%%%%%%%%%%%%%%%%%%%%%%%%%%%%%%

We investigate the angular resolution and distance determination of the BNS and NSBH sources at different redshifts for a single 3G detector as well as different networks of up to four 3G detectors. For a comparison of the design of CE with ET-D, seven different 3G network configurations are considered. Specifically, we consider

$\bullet$ {\bf 2CE} Two CE-like detectors, one in the US and one in Europe, at the same site and with same orientation as ET, but with an opening angle of $90^{\circ}$.

$\bullet$ {\bf CE-ETD} One CE detector in the US, and one ET detector in Europe with the ET-D sensitivity.

$\bullet$ {\bf 2ETD} Two ET-like detectors with the ET-D sensitivity, one in Europe and one in the US, at the same site and with the same orientation as CE, but with an opening angle of $60^{\circ}$.

$\bullet$ {\bf 3CE} Three CE-like detectors, one in the US and one in Europe, plus one in Australia.

$\bullet$ {\bf 3ETD} Three ET-like detectors with the ET-D sensitivity, one in the US and one in Europe, plus one in Australia.

$\bullet$ {\bf 3cETD} Three ET-like detectors with ET-D sensitivity, one in the US and one in Europe, plus one in China.

$\bullet$ {\bf 4ETD} Four ET-like detectors with ET-D sensitivity, one in the US, one in Europe, one in China and one in Australia.

The coordinates, orientations, and open angles of the detectors are listed in Table \ref{table1}. We stress that the coordinates and orientations do not represent the actual localizations of the potential detector, and we did not check for the optimal site localizations for 3G networks, which have been well studied in \cite{3G-best-position}. We use 15,000 simulated random BNS samples of {$m_{1,{\rm phys}}=m_{2,{\rm phys}}=1.4$ \mSun ~for each redshift values of $z\in [0.1,2]$ with 0.1 spacing, and for NSBH system of fixed mass pairs of $m_{1,{\rm phys}}=10$ \mSun ~and $m_{2,{\rm phys}}=1.4$ \mSun ~sampled at $z\in[0.1,2]$ as prescribed in Sec.~\ref{sec:Fisher}. %The sky directions are chosenly randomly from} {\r ra $\in[0,360]$ deg, cos(dec) $\in[-1,1]$ and binary orientations and initial phase}.
%The redshift values are distributed uniformly in the range of $z\in[0.1,1.0]$ with a 0.1 spacing.   }

\label{sec:Network}
\subsection{One detector}

%{\r [discuss the relation between angular resolution and distance determination somewhere]}
{
We consider ET-D for one-detector localization and distance determination of BNS and NSBH sources at different redshifts as it has the best single detector angular resolutions (Sec.~\ref{sec:One3G}).  The results for the BNS systems are shown in violin plots in Fig.~\ref{ft1} (upper panels), and for the NSBH systems in Fig.~\ref{ft1} (lower panels). We find that for one ET-D detector, the medium angular resolution at $z=0.1$ for the detected BNS and NSBH sources is around 100 deg$^2$ and the medium distance accuracy is about $20\%$.   This is already comparable to the localization of the first detected GW event GW150914 by the two Advanced LIGO detectors at O1 sensitivity \cite{gw150914}.  At $z=0.2$, the medium angular resolution is $ 1000~{\rm deg}^2$ and medium $\Delta d_{\rm L}/d_{\rm L}\sim 40\%$.

{We have observed a pronounced anti-correlation relation of the distance accuracy with the angular resolution for GW sources for all types of detectors.  For one ET-D detector, such  anti-correlation relation is shown in Fig.~\ref{fz3} for our BNS samples at $z=0.1$.  The angular resolution of more face-on sources (Fig.~\ref{fz3}, right panels, red and blue points) are overall better than more edge-on sources (Fig.~\ref{fz3}, right panels, green and yellow points) due to their larger SNRs. On the other hand, the GW polarization for edge-on systems is more linearized and can be better determined by one detector. This together with the time-dependent antenna beam-pattern function helps break its degeneracy with the distance determination.   Among those face-on sources,  however, the angular resolution is worse for those with a direction perpendicular to the detector plane despite of their larger SNRs.  At these directions, the influence of the Earth's rotation is minimum, and therefore the single-detector angular resolution is the worst.    Similar conclusion holds for the NSBH sources.   %From Fig.~\ref{fz3}, for these sources, we also find a pronounced anti-correlation of the angular resolution and SNR. Comparing two panels of this figure, we arrive at the conclusion that for the single 3G detector, like ET,
Therefore, for a single ET-D like detector, the face-on sources from the directions parallel to the detector plane have the best angular resolution, while the edge-on sources have overall better distance determination.}

%Applying the same analysis to the ideal experiment, we investigate the localizations of BNSs and NSBHs at different redshifts, which are plotted in Fig.~\ref{ft2}. We find that by an individual ideal detector, even for the sources at $z=2$, we have $\Delta\Omega_s<100$deg$^2$ for all the samples, and $\Delta d_{\rm L}/d_{\rm L}<0.1$ for most samples. In particular, for the BNSs at $z=0.1$, we have $\Delta\Omega_s \sim 1$deg$^2$ and $\Delta d_{\rm L}/d_{\rm L}\sim0.01$, and for NSBHs at this redshift, we have $\Delta\Omega_s \sim 0.1$deg$^2$ and $\Delta d_{\rm L}/d_{\rm L}\sim0.005$, which are greatly helpful to identify their host galaxies.  Due to the effect of the Earth's rotation, an individual ideal detector is good enough for the localization of compact binary systems with small masses. However, it is argued that for the ground-based detector, it might be difficult to reduce the noise at $f\sim 1$ Hz to this low level, but it is relatively easier to reduce the noise at $f<5$ Hz \cite{3G-lowfreq,ce}.  As a conservative consideration, {\b we} consider the ideal detector, but adopt the low-frequency cutoff at $5$ Hz, instead of $1$ Hz, the violin plots for $\Delta\Omega_s$ and $\Delta d_{\rm L}/d_{\rm L}$ with respective to redshift are plotted in Fig.~\ref{ft2}.
{
In comparison, we replace ET-D with the ideal detector that has a better low-frequency sensitivity than ET-D (Fig.~\ref{f0}) and perform the same calculations also for $15,000$ random samples of BNSs and NSBHs. We found that for both BNSs and NSBHs, we have an order of magnitude improvement in the angular resolution and a factor of a few improvement in distance determination.  Specifically,  at $z=0.1$  the medium angular resolution is $\sim 10$ deg$^2$ and medium $\Delta d_{\rm L}/d_{\rm L}\sim 4 $\%, while at $z=0.3$ the medium $\Delta \Omega_s\sim 100$ deg$^2$ and medium $\Delta d_{\rm L}/d_{\rm L}\sim 15$\%. For both ET-D and the ideal detector, we find that the angular resolutions and distance determinations of NSBHs are slightly worse than that of BNSs, since the effect of the Earth's rotation is weaker for NSBHs. %Such an ideal detector can extend the accuracy of the ET-D detector from $z=0.1$ to $z=0.2$.
}
%%%%%%%%%%%%%%%%%%%%%%%%%%%%%%%%%%%%%%%%%  figure 1, figure 1, figure1%%%%%%%%%%%%%%%%%%%%%%%%%%%%%%%%%%%%%%%%%%%%%%%%%%
%\begin{figure}
%\begin{center}
%\centerline{\includegraphics[width=15cm]{figs2/single/violin/ideal_nsns_nsbh.png}}
%\end{center}\caption{Same with Fig.~\ref{ft1}, but ET-D noise is replaced by the ideal noise with low-frequency cutoff $f_{\rm low}=1$ Hz, and the redshift values are in the range $z\in[0.1,2.0]$.}\label{ft2}
%\end{figure}
%%%%%%%%%%%%%%%%%%%%%%%%%%%%%%%%%%%%%%%%  figure 1, figure 1, figure1%%%%%%%%%%%%%%%%%%%%%%%%%%%%%%%%%%%%%%%%%%%%%%%%%%

%\section{Localization of GW sources by the detector networks}

%As well known, for the ground-based GW detectors, a network with large baselines can facilitate the localization of GW sources. Until now, the Europe proposed ET and US proposed CE have been widely studied in the literature. In the section, employing the Fisher matrix, we investigate the localization abilities of GW sources for third-generation network including two or several ET-like and/or CE-detectors. In particular, we shall focus on the difference between ET-like and CE-like detectors in the networks.

\subsection{Network of two detectors}
\label{sec:2Network}

%Here, let us investigate the localization capabilities of the 3G detector networks.

%{\r [how do you decide on the orientation of CE in Europe? describe it somewhere? ]}

%%%%%%%%%%%%%%%%%%%%%%%%%%%%%%%%%%%%%%%%%  figure 1, figure 1, figure1%%%%%%%%%%%%%%%%%%%%%%%%%%%%%%%%%%%%%%%%%%%%%%%%%%
\begin{figure}
\begin{center}
\centerline{\includegraphics[width=15cm]{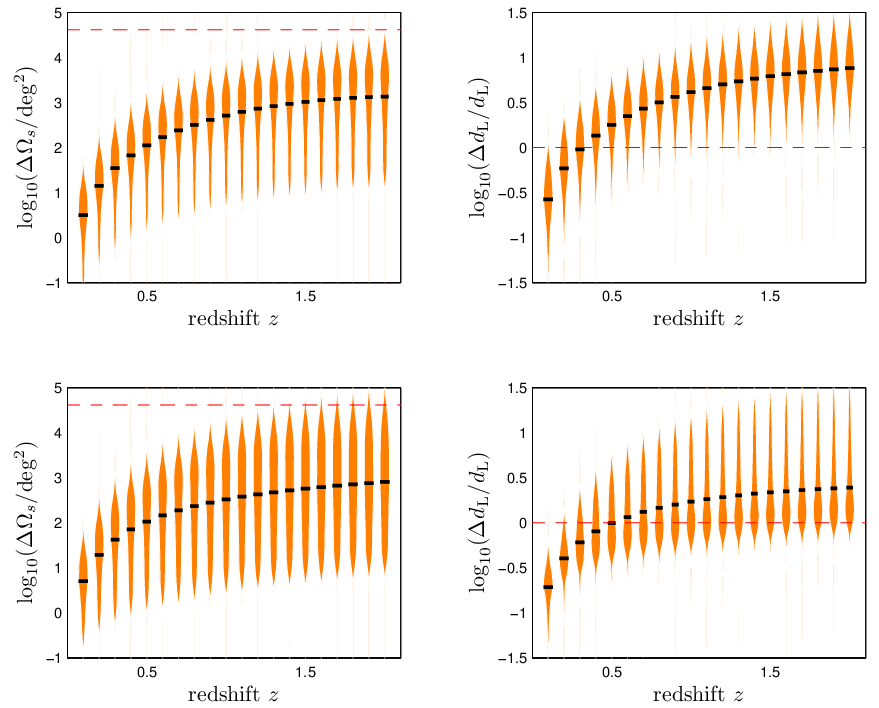}}
\end{center}\caption{Same as Fig.~\ref{ft1}, but for {\b the 2CE network}.}\label{ft3}
\end{figure}

{The result for the 2CE network can be found in Fig.~\ref{ft3}.
At $z=0.1$, the medium $\Delta\Omega_s$ is around a few square-degrees, and medium $\Delta d_{\rm L}/d_{\rm L}$ is a few percent for BNSs (Fig.~\ref{ft3}, upper panels) and similarly for NSBHs (Fig.~\ref{ft3}, lower panels). The medium angular resolution at $z=0.2$ is about 20 deg$^2$ and medium distance accuracy  $\sim 50\%$. On the other hand, there is a small but noticeable fraction of BNS and NSBH sources that can be localized to sub-square degree accuracy at $z<0.3$, and with distance accuracy to be less than $10\%$ at $z=0.1$.  Even for the sources at $z=2$, there is still a small but noticeable fraction of sources that can have angular resolution to be within $\Delta\Omega_s\sim 10$ deg$^2$.  As the effect of the Earth rotation is nearly negligible for the CE detector, these are mainly due to the much larger SNRs from the superior sensitivity of the CE detectors, especially for those face-on binaries. {Note also,  the fraction of BNSs detected by the two CE detectors is also much more than the two ET-D detectors} {(Fig.~\ref{fy1})}.    %Clearly a detector of more than 2 detectors are desirable in order to better localise sources up to $z>0.5$.
}

%{\r [Add Omega-dD/D figure here and discuss ]}

%{\b However, this figure also shows that the spreads of the distributions of both $\Omega_s$ and $\Delta d_{\rm L}/d_{\rm L}$ are too wide. For instance, for the BNSs at $z=2$, we have $\Omega_s\in(10,4\times10^4)$deg$^2$, and the medial value is $\Omega_s\sim 10^3$deg$^2$, which indicates that the 2-detector networks consisting of only CE-like detectors cannot well localize most GW sources at this redshift. So, we should add new detector in the networks, or consider the detectors of other types.

%Now, let us turn to the network including ET or ET-like detectors.
{The medium angular resolution will be greatly improved for sources at high-$z$ when the 2-detector network include at least one ET-D due to its low-frequency sensitivities as discussed previously. In comparison with the 2CE network at the same locations, we find a medium angular resolution of a few degrees at $z=0.1$, 15 deg$^2$ at $z \sim 0.2$, and $\sim 100 $ deg$^2$ at $z=2$ for the CE-ETD network.  The improvement in distance accuracy is also a factor of a few better than the 2CE network. However, the fraction {of sources with superior localization observed with the 2CE network is reduced}.

{If both detectors are of the ET-D type, nearly all the BNS sources at $z<0.2$ can be localized within 10 deg$^2$ (Fig.~\ref{ft5}).  The medium angular resolution for the BNS sources is  $\lesssim 1$ deg$^2$ at $z=0.1$ (Fig.~\ref{ft5}, left panels), and 10 deg$^2$ at $z=0.4$.  The medium angular resolution for NSBH sources is slightly worse. It is around 10 deg$^2$ at $z=0.3$.  The 10\% medium distance uncertainty can be achieved at $z=0.3$ for both the BNS and NSBH sources (Fig~\ref{ft5}, right panels).}

\begin{figure}
\begin{center}
\centerline{\includegraphics[width=15cm]{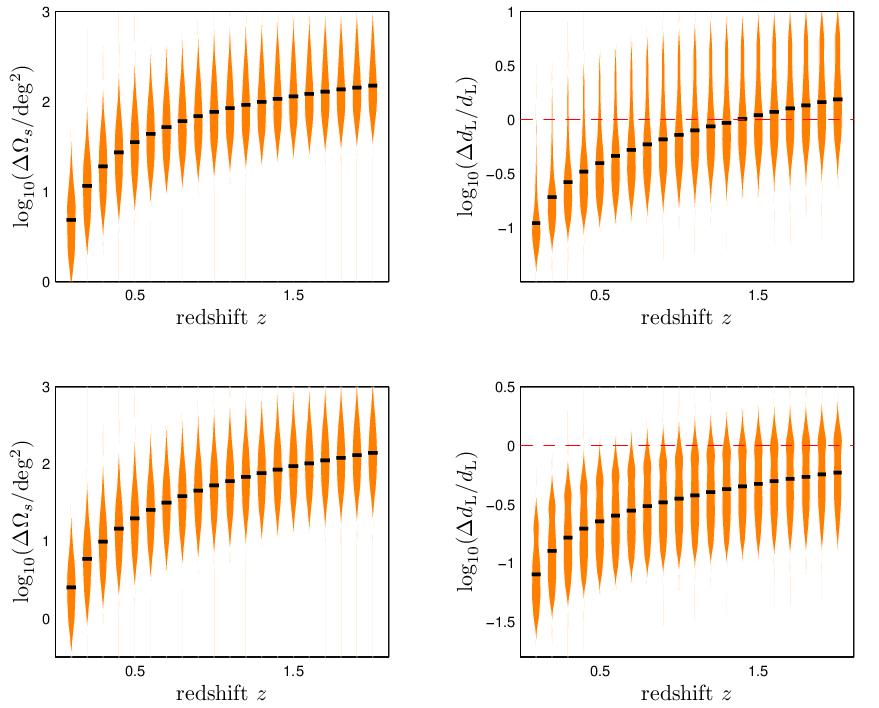}}
\end{center}\caption{Same as Fig.~\ref{ft3}, but for {\b the CE-ETD network}.}\label{ft4}
\end{figure}
%%%%%%%%%%%%%%%%%%%%%%%%%%%%%%%%%%%%%%%%  figure 1, figure 1, figure1%%%%%%%%%%%%%%%%%%%%%%%%%%%%%%%%%%%%%%%%%%%%%%%%%%

In Fig.~\ref{fz4}, we present the 2-dimensional $\Delta\Omega_s$-$\Delta d_{\rm L}/d_{\rm L}$ scatter plot of BNS samples at $z=0.1$ detected by the 2ETD network. Similar to the single-detector case in Fig.~\ref{fz3}, there is a pronounced anti-correlation between the angular resolution and distance accuracy.  Similarly, the face-on binaries have overall better angular resolution and SNRs, but worse distance accuracy.  In the meanwhile, the edge-on binaries have a better distance determination, but worse angular resolution and SNRs.  We can also confirm that the anti-correlation relation holds for both BNSs and NSBHs at any given redshift.  {Note the relation of angular resolution and the GW direction relative to the detector plane is no longer obvious as the baseline effect dominates for a network of detectors. }

% We further consider a 2-detector network consisting of two ET experiments: One is at Europe and the other is at US. Both of them have ET-D noise. The corresponding violin plots are given in Fig.~\ref{ft5}. In comparison with the previous results, we find that due to the special configuration of ET, the values of $\Delta\Omega_s$ and $\Delta d_{\rm L}/d_{\rm L}$ become smaller, and the spreads of their distribution become even narrower: For the sources at $z=0.1$, we have $\Delta\Omega_s\in(0.15,3){\rm deg}^2$ and $\Delta d_{\rm L}/d_{\rm L}\in(0.015,0.5)$ for BNSs, $\Delta\Omega_s\in(0.6,10){\rm deg}^2$ and $\Delta d_{\rm L}/d_{\rm L}\in(0.025,0.3)$ for NSBHs. For the sources at $z=1$, they are $\Delta\Omega_s\in(10,100){\rm deg}^2$ and $\Delta d_{\rm L}/d_{\rm L}\in(0.1,1)$ for BNSs, $\Delta\Omega_s\in(10,200){\rm deg}^2$ and $\Delta d_{\rm L}/d_{\rm L}\in(0.05,1)$ for NSBHs. For the remote sources at $z=2$, both BNSs and NSBHs have $\Delta\Omega_s\in(20,200)$deg$^2$, and nearly half of them have $\Delta d_{\rm L}/d_{\rm L}<1$. In addition, from Fig.~\ref{ft5} we also find that the distributions of $\Delta\Omega_s$ nearly keep constant once $z>1$, which means that by this network, the localization of GW sources are quite accuracy even they are localized at high redshift.

In summary, we can conclude that (1) For the network consisting of only CE-like detectors, accurate localization can be achieved mainly for nearby sources with $100$ deg$^2$ medium angular resolution achieved at $z=0.5$. {However, due to the superior sensitivity of the CE detector, the total number of sources with superior localization at low-redshift might be comparable to that of the 2ETD network. (2) For CE-ETD and 2ETD networks, the accurate localization can be greatly extended to high redshift $z\sim 2$, where we have $\Delta\Omega_s\sim 100$ deg$^2$ for most BNSs and NSBHs. In particular for the nearest sources detected by 2ETD network at $z=0.1$, the localization can be as accurate as $\Delta\Omega_s\sim 1$ deg$^2$ and $\Delta d_{\rm L}/d_{\rm L}\sim 5\%$ for most sources. (3) We observe again that for binaries coalescences at a fixed redshift, the face-on binaries {can be detected with a better angular resolution, while the edge-on binaries with a better luminosity distance accuracy. }

%(3) Relative to the individual detector, the effect of the Earth's rotation is much smaller for the detector networks.

%The histograms are plotted in Fig.~\ref{fd2}, where we find that the accuracy of source localization are quite similar with the network consisting of ET and CE, but the distribution is much more narrow.

%we plot the same histograms by considering the detector network including the ET in Europe and a similar detector in US, which shows the similar results with Fig.~\ref{fd1}. Both results show that the effect of the Earth's rotation is not significant for the third-generation networks, if the noise levels of them are similar with ET-B, ET-D and/or CE.

%\begin{figure}
%\begin{center}
%\centerline{\includegraphics[width=15cm]{figs2/network/hist/ETDETD/NSNS_NSBH_4_SNR8_2.png}}
%\end{center}\caption{Same with Fig.~\ref{fa1}, but here we consider the network of GW detectors including the ET in Europe and an ET-like detector in US.}\label{fd2}
%\end{figure}

%%%%%%%%%%%%%%%%%%%%%%%%%%%%%%%%%%%%%%%%%  figure 1, figure 1, figure1%%%%%%%%%%%%%%%%%%%%%%%%%%%%%%%%%%%%%%%%%%%%%%%%%%
\begin{figure}
\begin{center}
\centerline{\includegraphics[width=15cm]{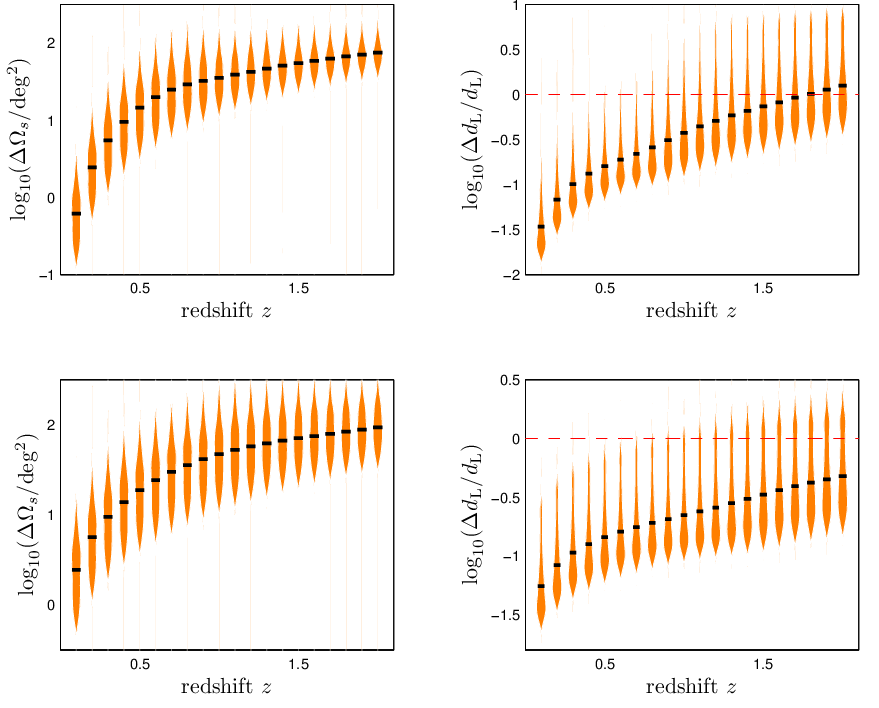}}
\end{center}\caption{Same as Fig.~\ref{ft3}, but for {\b the 2ETD network}.}\label{ft5}
\end{figure}
%%%%%%%%%%%%%%%%%%%%%%%%%%%%%%%%%%%%%%%%  figure 1, figure 1, figure1%%%%%%%%%%%%%%%%%%%%%%%%%%%%%%%%%%%%%%%%%%%%%%%%%%

%%%%%%%%%%%%%%%%%%%%%%%%%%%%%%%%%%%%%%%%%  figure 1, figure 1, figure1%%%%%%%%%%%%%%%%%%%%%%%%%%%%%%%%%%%%%%%%%%%%%%%%%%
\begin{figure}
\begin{center}
\centerline{\includegraphics[width=15cm,height=10cm]{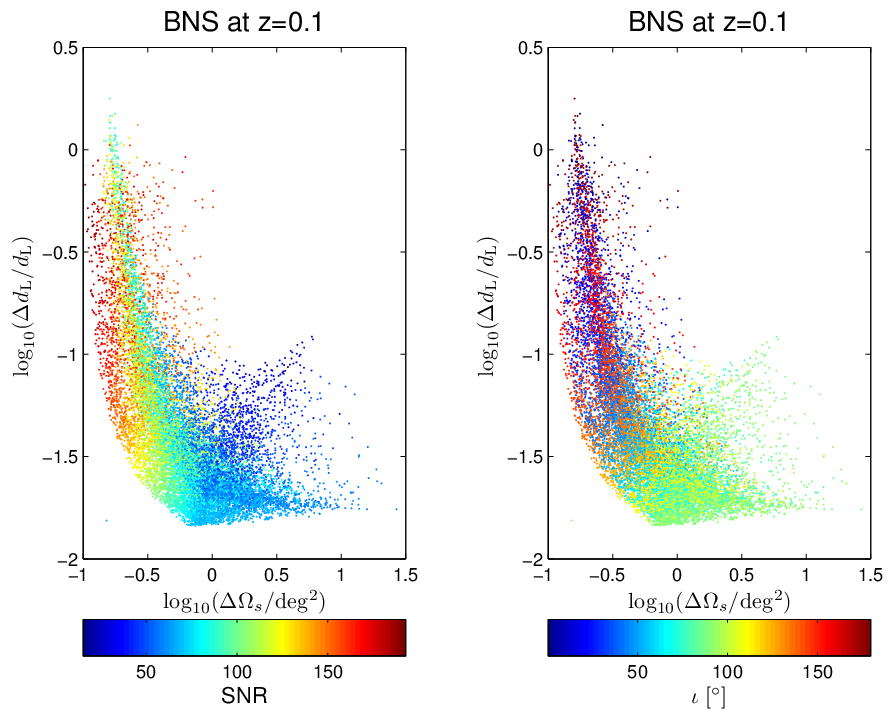}}
\end{center}\caption{Same as Fig.~\ref{fz3}, but a single ET-D detector is replaced by {\b the 2ETD network}.}\label{fz4}
\end{figure}

\subsection{Networks of three and four detectors}
\label{sec:3Network}
%The conception of 3G experiments is still under consideration, it is interesting to investigate the capabilities of the potential networks, which contains several 3G detectors.
The benefits of building a large network of GW detectors including Australia and China have been widely discussed \cite{blair2015,Howell2017,vilta}. The advantages of multi-detector networks for the localization of GW sources have been studied in a previous work \cite{vilta} for BBHs with component masses larger than 12 \mSun.  In this paper, we focus instead on binaries with small masses, i.e. BNSs and NSBHs that have significantly long-duration signals in the frequency band of 3G detectors.

The angular resolution and distance accuracy for the 3CE network are shown in Fig.~\ref{ft6} for both BNSs and NSBHs in the redshift range $z\in[0.1,2]$. Compared with the results of 2CE network in Fig.~\ref{ft3}, we find that a 3CE network can greatly improve the localization accuracy for binary coalescences at any redshift. At $z=0.1$, we have $\Delta\Omega_s\in(0.005,0.6)~$deg$^2$ and $\Delta d_{\rm L}/d_{\rm L}\in(0.006,0.25)$ for BNSs, $\Delta\Omega_s\in(0.02,0.5)~$deg$^2$ and $\Delta d_{\rm L}/d_{\rm L}\in(0.01,0.25)$ for NSBHs. At $z=1$, we have $\Delta\Omega_s\in(0.3,30)~$deg$^2$ and $\Delta d_{\rm L}/d_{\rm L}\in(0.03,1)$ for BNSs, $\Delta\Omega_s\in(0.3,50)~$deg$^2$ and $\Delta d_{\rm L}/d_{\rm L}\in(0.03,1)$ for NSBHs. At $z=2$, we still have $\Delta\Omega_s\in(2,100)~$deg$^2$ for both BNSs and NSBHs, while the medium $\Delta d_{\rm L}/d_{\rm L}=30\%$ for BNSs and medium $\Delta d_{\rm L}/d_{\rm L}=20\%$ for NSBHs.

Fig.~\ref{ft7} shows the results for the 3ETD network. Compared with the 3CE network, we find that this network gives similar, but slightly narrower distributions, for the localization of GW sources at all redshifts. In Fig.~\ref{fy1}, we plot the fraction of detectable BNS and NSBH sources as a function of redshift $z$ for various networks. Thanks to the lower noise level, the 3CE network has much better detection rates at high redshift $z>0.5$ than all other detector networks in consideration. For instance, at $z=2$, the 3CE network can detect $90\%$ BNSs and $98\%$ NSBHs, while the 3ETD network detects only $38\%$ BNSs and $74\%$ NSBHs.

%%%%%%%%%%%%%%%%%%%%%%%%%%%%%%%%%%%%%%%%%  figure 1, figure 1, figure1%%%%%%%%%%%%%%%%%%%%%%%%%%%%%%%%%%%%%%%%%%%%%%%%%%
\begin{figure}
\begin{center}
\centerline{\includegraphics[width=15cm]{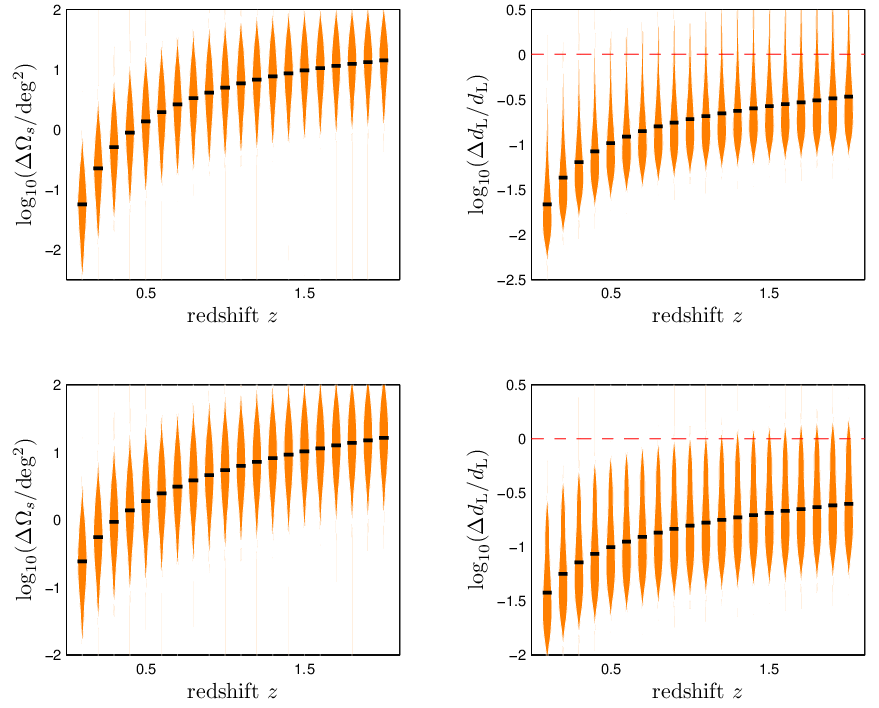}}
\end{center}\caption{Same as Fig.~\ref{ft3}, but for {\b the 3CE network}.}\label{ft6}
\end{figure}
%%%%%%%%%%%%%%%%%%%%%%%%%%%%%%%%%%%%%%%%  figure 1, figure 1, figure1%%%%%%%%%%%%%%%%%%%%%%%%%%%%%%%%%%%%%%%%%%%%%%%%%%

%%%%%%%%%%%%%%%%%%%%%%%%%%%%%%%%%%%%%%%%%  figure 1, figure 1, figure1%%%%%%%%%%%%%%%%%%%%%%%%%%%%%%%%%%%%%%%%%%%%%%%%%%
\begin{figure}
\begin{center}
\centerline{\includegraphics[width=15cm]{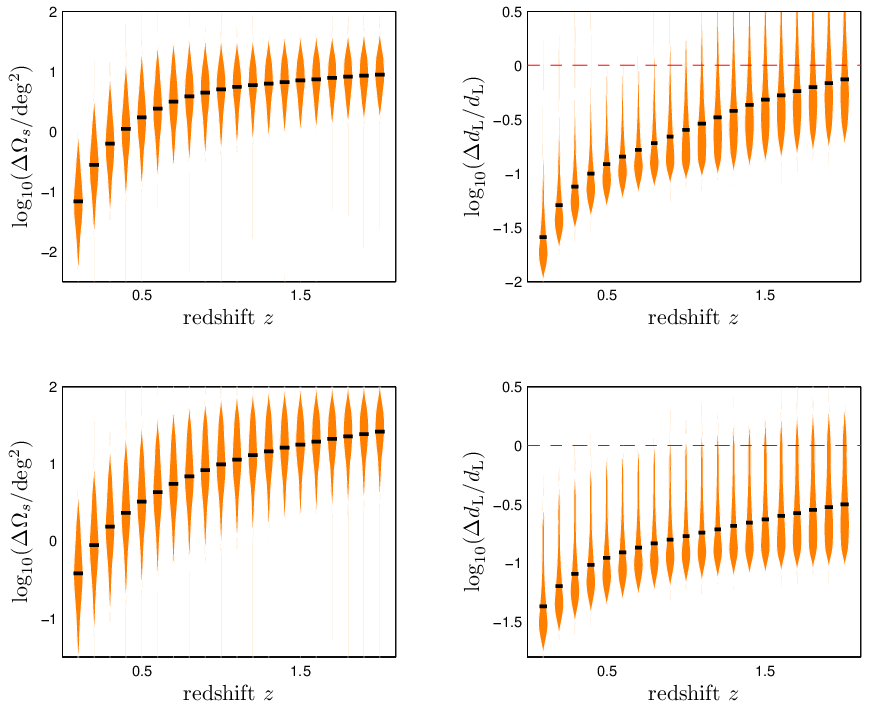}}
\end{center}\caption{Same as Fig.~\ref{ft3}, but for {\b the 3ETD network}.}\label{ft7}
\end{figure}
%%%%%%%%%%%%%%%%%%%%%%%%%%%%%%%%%%%%%%%%  figure 1, figure 1, figure1%%%%%%%%%%%%%%%%%%%%%%%%%%%%%%%%%%%%%%%%%%%%%%%%%%

%%%%%%%%%%%%%%%%%%%%%%%%%%%%%%%%%%%%%%%%%  figure 1, figure 1, figure1%%%%%%%%%%%%%%%%%%%%%%%%%%%%%%%%%%%%%%%%%%%%%%%%%%
\begin{figure}
\begin{center}
\centerline{\includegraphics[width=15cm]{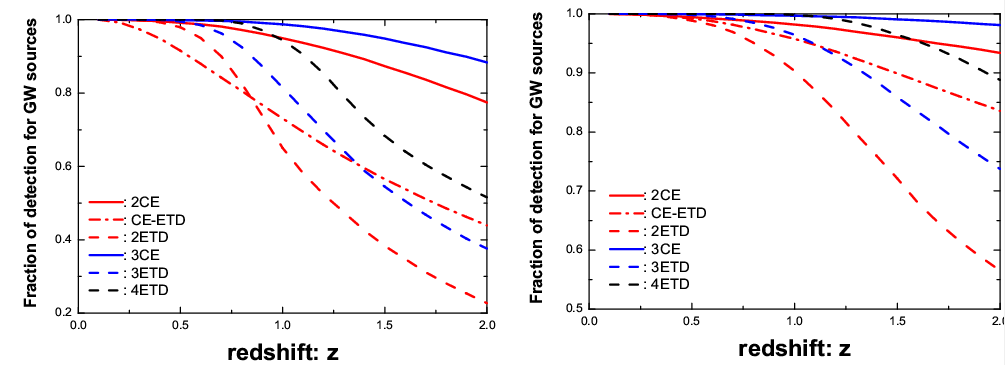}}
\end{center}\caption{For various detector networks, the fraction of detectable GW sources (i.e. $\rho>8$) as a function of redshift $z$. The left panel shows the results of BNSs and the right one is for NSBHs. }\label{fy1}
\end{figure}
%%%%%%%%%%%%%%%%%%%%%%%%%%%%%%%%%%%%%%%%  figure 1, figure 1, figure1%%%%%%%%%%%%%%%%%%%%%%%%%%%%%%%%%%%%%%%%%%%%%%%%%%

%%%%%%%%%%%%%%%%%%%%%%%%%%%%%%%%%%%%%%%%%  figure 1, figure 1, figure1%%%%%%%%%%%%%%%%%%%%%%%%%%%%%%%%%%%%%%%%%%%%%%%%%%
\begin{figure}
\begin{center}
\centerline{\includegraphics[width=15cm]{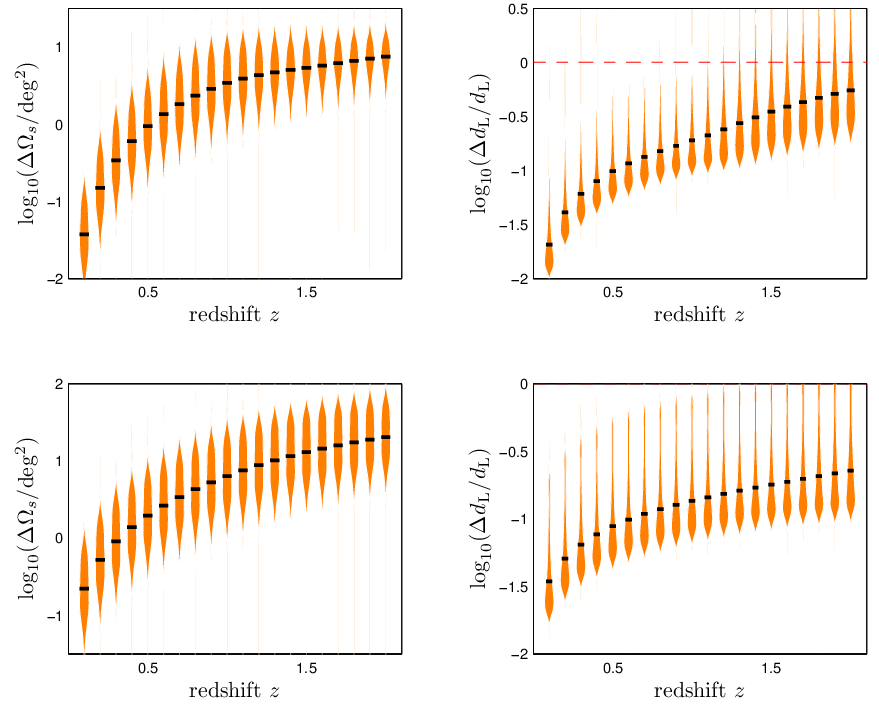}}
\end{center}\caption{Same as Fig.~\ref{ft3}, but for {\b the 4ETD network}.}\label{ft8}
\end{figure}
%%%%%%%%%%%%%%%%%%%%%%%%%%%%%%%%%%%%%%%%  figure 1, figure 1, figure1%%%%%%%%%%%%%%%%%%%%%%%%%%%%%%%%%%%%%%%%%%%%%%%%%%

%%%%%%%%%%%%%%%%%%%%%%%%%%%%%%%%%%%%%%%%%  figure 1, figure 1, figure1%%%%%%%%%%%%%%%%%%%%%%%%%%%%%%%%%%%%%%%%%%%%%%%%%%
\begin{figure}
\begin{center}
\centerline{\includegraphics[width=15cm,height=10cm]{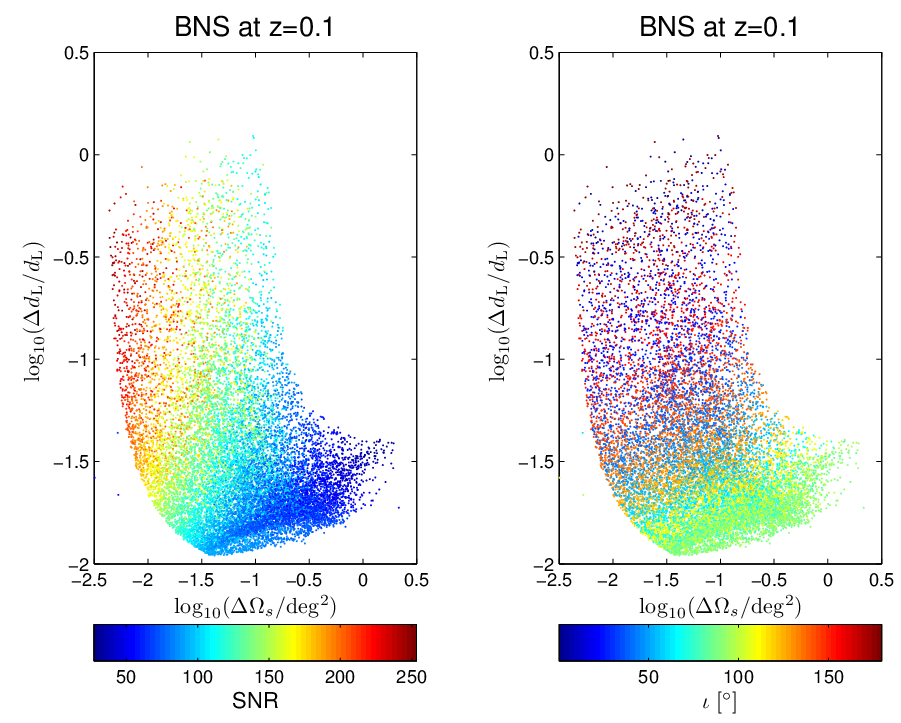}}
\end{center}\caption{Same as Fig.~\ref{fz3}, but the single ET-D detector is replaced by {\b the 3ETD network}.}\label{fz5}
\end{figure}
%%%%%%%%%%%%%%%%%%%%%%%%%%%%%%%%%%%%%%%%  figure 1, figure 1, figure1%%%%%%%%%%%%%%%%%%%%%%%%%%%%%%%%%%%%%%%%%%%%%%%%%%

%%%%%%%%%%%%%%%%%%%%%%%%%%%%%%%%%%%%%%%%%  figure 1, figure 1, figure1%%%%%%%%%%%%%%%%%%%%%%%%%%%%%%%%%%%%%%%%%%%%%%%%%%
\begin{figure}
\begin{center}
\centerline{\includegraphics[width=15cm,height=10cm]{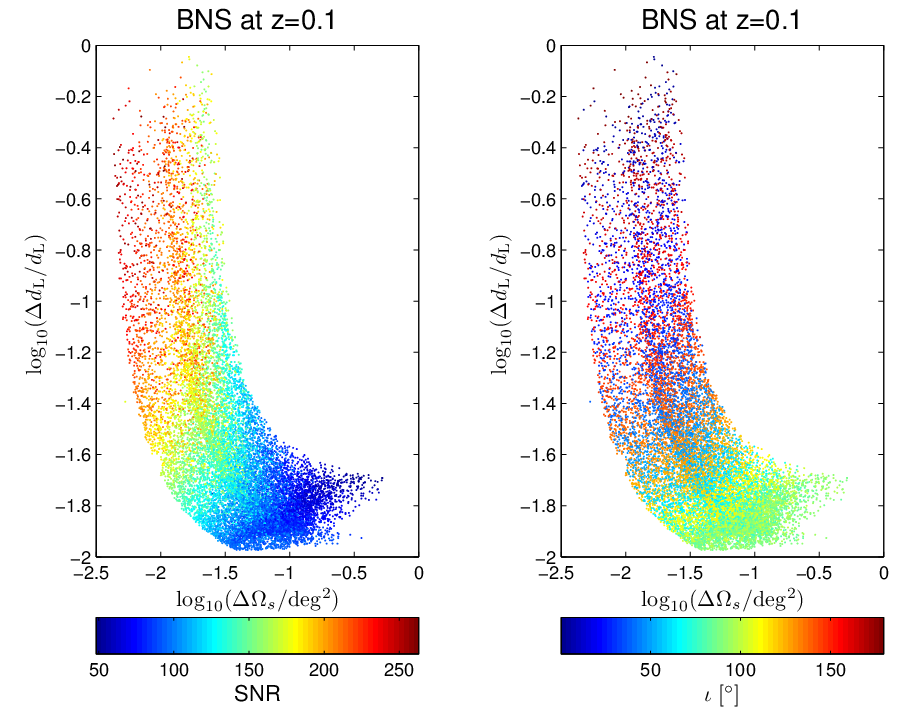}}
\end{center}\caption{Same as Fig.~\ref{fz3}, but the single ET-D detector is replaced by {\b the 4ETD network}.}\label{fz6}
\end{figure}
%%%%%%%%%%%%%%%%%%%%%%%%%%%%%%%%%%%%%%%%  figure 1, figure 1, figure1%%%%%%%%%%%%%%%%%%%%%%%%%%%%%%%%%%%%%%%%%%%%%%%%%%

For the 4ETD network, we find that the localization accuracies of binary coalescences are slightly better than that for the 3-detector networks. For BNSs, the angular resolution $\Delta\Omega_s$ can be improved by a factor $\sim 2$, and $\Delta d_{\rm L}/d_{\rm L}$ shows no significant improvement. For the NSBHs, we find the medium values of both $\Delta\Omega_s$ and $\Delta d_{\rm L}/d_{\rm L}$ have no significant improvement, from that of the 3-detector networks at any redshift. However, the spreads of the distributions of $\Delta\Omega_s$ for 4ETD network are much narrows than the 3-detector networks. For instance, for the BNSs at $z<0.1$, we find most sources can have a localization accuracy of $\Delta\Omega\in(0.01,0.1)$~deg$^2$, which is significantly smaller to directly identify their host galaxies. Even for the BNSs at $z=2$, we also have $\Delta\Omega\in(2,10)$~deg$^2$, which is helpful for follow-up observations on their EM counterparts.

From Fig.~\ref{fz5} and Fig.~\ref{fz6}, we again find a significant anti-correlation between $\Delta\Omega_s$ and $\Delta d_{\rm L}/d_{\rm L}$ for BNSs at $z=0.1$, similar to the cases of a single ET-D detector and of the 2ETD network. The face-on GW sources have a better angular resolution, while the edge-on sources have a better distance determination. For 3G GW detectors, regardless an individual detector or detector networks, it seems impossible to have both the best angular resolution and the best distance determination at the same time. However, as long as there are more than one detector,  there is always a positive correlation between the angular resolution and SNR.

%However, different from the case of a single ET-D detector, for the face-on sources observed by 3G detector networks, we find a pronounced positive correlation between SNR and the angular resolution, and the sources with the largest SNRs have the best angular resolution.

%These results demonstrate that, the network consisting of three three-generation detectors in Europe, US and Australia, is good enough for the localization of GW sources at any redshift.

%[ The only "anomaly" in single detector are those with direction perpendicular to the plane of the detectors, otherwise, the angular resolution is always better with high SNRs.]

%Change to:

%However, as long as there are more than one detector,  there is always a positive correlation between the angular resolution and SNR?

\subsection{Effect of a larger network}

The angular resolution of a GW source detected by detector networks depends not only on the area of the errors $\Delta\Omega_s$, but also on the 2-dimensional shapes of error ellipses in the two sky directions. For the three 2-detector networks considered, we show in Fig.~\ref{fx1} the sky maps of the error ellipses of angular parameters derived for the BNSs at the redshift $z=2$. In the left panels, for each sample, we adopt $\iota=0^{\circ}$, i.e., the face-on sources, while in right panels $\iota=45^{\circ}$ for comparison. For face-on sources, which have the most accurate medium angular resolutions among all the sources at same redshift, we find that 2CE network gives the best localization accuracy. For the networks consisting of one or two ET-Ds, the angular resolution of the face-on sources are slightly worse, due to lower SNRs. These results are consistent with those in violin plots in Figs.~\ref{ft3}, \ref{ft4} and \ref{ft5}. However, for sources with general inclination angles, from right panels of Fig.~\ref{fx1}, we find that 2ETD and CE-ETD networks always achieve much more accurate medium angular resolution.
For these 2-detector networks, we also find that in most region of sky, the error ellipses are string-shaped as it is mostly determined by the arrival time delay of GWs between two detectors. This is similar to the localization of the GW events GW150914, as well as GW151226 and GW170104, observed by two Advanced LIGO detectors \cite{gw150914,gw151226,gw170104}. The shapes of the error ellipses strongly depend on the positions of the GW sources in the sky, as well as locations of detectors. However, the far more superior single-detector angular resolution of ET-D helps remove some of the extremely long ellipses in the sky map of the 2ETD network, which is similar to the localization of the GW event GW170814, observed by the collaboration of two Advanced LIGO detectors and one Advanced Virgo detector \cite{gw170814}. These features also hold for the NSBH sources.

\begin{figure}
\begin{center}
\centerline{\includegraphics[width=15cm]{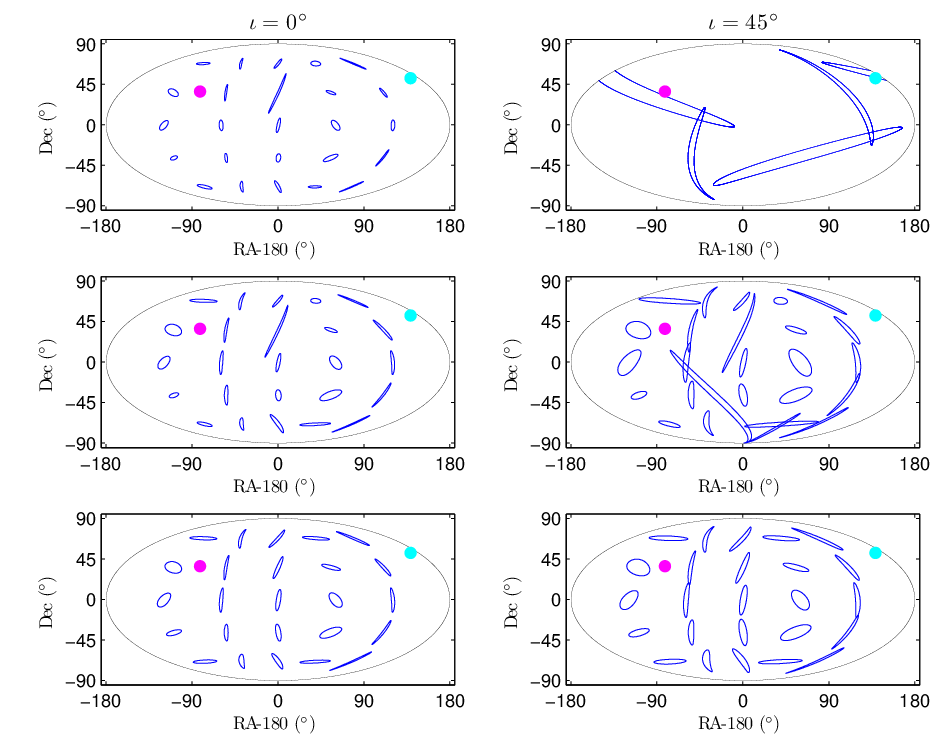}}
\end{center}\caption{For the BNSs at $z=2$, in equatorial coordinate system, this figure shows all-sky map of error ellipses of angular parameters for 2-detector networks of {\b 2CE} (upper panels), {\b CE-ETD} (middle panels), {\b 2ETD} (lower panels).  The left panels show the results of face-on sources, and the right panels show that with the inclination angle $\iota=45^{\circ}$. In each panel, the magenta dot denotes the location of European detector at $t=0$ in equatorial coordinate, the cyan dot denotes that of American detector. }\label{fx1}
\end{figure}

\begin{figure}
\begin{center}
\centerline{\includegraphics[width=15cm]{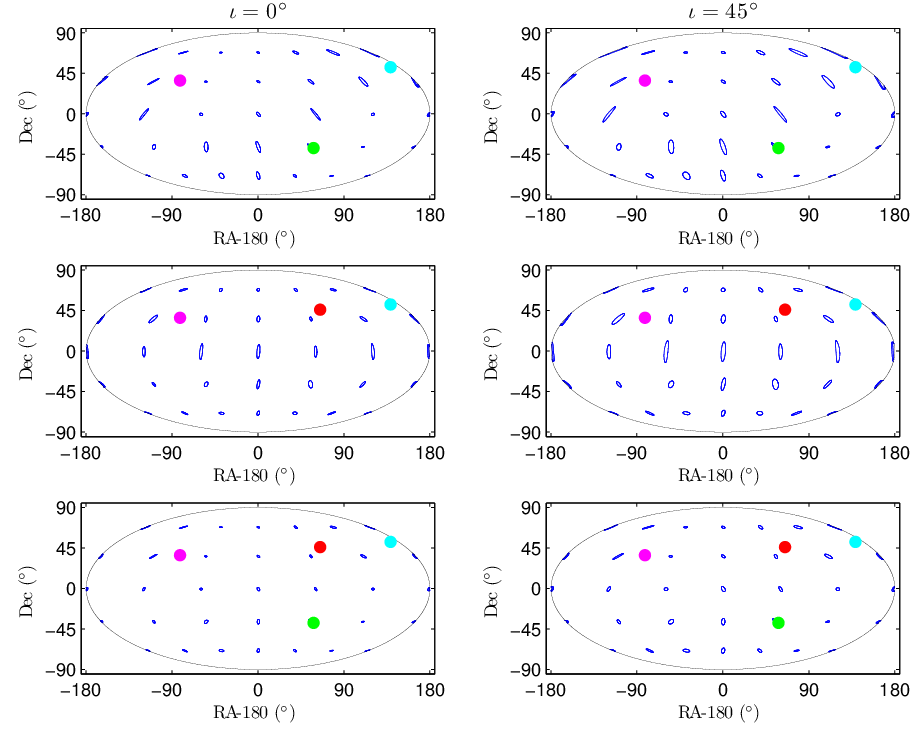}}
\end{center}\caption{Same as Fig.~\ref{fx1}, but for the results of detector networks {\b 3ETD} (upper panels), {\b 3cETD} (middle panels) and {\b 4ETD} (lower panels). In each panel, the magenta dot denotes the location of European detector at $t=0$ in equatorial coordinate, the cyan dot denotes that of American detector, the green dot denotes that of Australian detector, and the red dot denotes that of Chinese detector.}\label{fx2}
\end{figure}

For the BNS sources at $z=2$, we show in Fig.~\ref{fx2} the sky maps of the error ellipses of sky directions for 3-detector and 4-detector networks. Comparing with the results of 2-detector networks, we find that the error ellipses are less elongated as expected from the contribution from more than one set of long baselines between detectors for these multi-detector networks. The 3ETD network and 3cETD network show similar results for binaries with different inclination angles. One noticeable difference is that, the orientation of the error ellipses at a given sky direction is different for different networks, which depends on the detector's sites on the Earth. In addition, for the 4-detector network, consisting of detectors in Europe, US, Australia and China, we find the localization of detector network becomes isotropic for binary coalescences from all sky directions. This means that increasing the detector number can help significantly remove observational blind regions in the sky.

\section{Implication for cosmology}
\label{sec:Cosmology}

As shown in previous sections, from the GW signal itself one can measure the luminosity distance of the binary coalescences independently without having to rely on a cosmic distance ladder. If the redshfit of the GW sources could also be measured by other means, such as by observing their EM counterpart or identifying the host galaxies, these GW sources can be used as \emph{standard sirens}, and $d_{\rm L}-z$ relation can be used to study the evolution history of the Universe \cite{schutz1986}. With the 2G detector network, the GW standard sirens at low redshift have been applied to measure the Hubble constant \cite{GW-cos1}. If the high redshift GW signals can be observed by 3G ground-based detectors \cite{sathya2009,zhao-2011,cai}, space-based LISA \cite{GW-cos2,GW-cos3,GW-cos4,GW-cos5}, BBO or DECIGO \cite{GW-cos6}, these sources are useful to determine the EoS of dark energy. In this section, based on the GW source localization of 3G detector networks, we consider the constraint on the Hubble constant $H_0$, the deceleration parameter $q_0$, and the EoS of cosmic dark energy by the inspirling BNSs and NSBHs.

\subsection{Determination of Hubble constant}

In the Friedmann-Lema\^{\i}tre-Robertson-Walker (FLRW) universe, if $z\ll 1$, the luminosity distance and the redshift satisfy the Hubble's law, which reads $d_{\rm L}=z/H_0$, where $H_0$ is the Hubble constant. The current Hubble constant is $H_0=67.8\pm 0.9~{\rm km~s}^{-1}{\rm~Mpc}^{-1}$, derived from the cosmic microwave background (CMB) radiation data \cite{planck}. Meanwhile, the large-scale structure observations inferred a different value $H_0=73.24\pm 1.74~{\rm km~s}^{-1}{\rm~Mpc}^{-1}$ \cite{zhaogongbo}, which conflicts with the CMB result at more than 3$\sigma$ confidence level.
The tension of measured $H_0$ value derived from different observations is one of the crucial puzzles of modern cosmology. On the other hand, observing GWs from binary coalescences at low-redshift can help determine the sources' luminosity distance at reasonably high precisions. The redshifts of these events can possibly be obtained with many methods. One possible way is to observe their EM counterparts, for instance, BNS and NSBH are hypothesized to be accompanied by short-hard GRBs or kilonova emissions, or to identify their host galaxies through their excellent position resolutions. Some other methods are also proposed, such as by detecting the contributions to the GW phases caused by tidal effects \cite{tidal}, by assuming a narrow mass distribution of NSs \cite{taylor}, by using the signature encoded in the post-merge signal \cite{post-merger}, or by directly observing the cosmological phase drift of a single GW source \cite{low-f-effect}. In this paper, we bypass the details and assume that one of the methods can be used to obtain the redshift information. {\r{For instance, combining with the observations of its EM counterpart, the GW data from the recently detected BNS event GW170817 has already been used to place an interesting constraint on $H_0$ with $H_0=70.0^{+12.0}_{-8.0}~{\rm km~s}^{-1}{\rm~Mpc}^{-1}$ \cite{gw170817-hubble}, which is consistent with the results derived from both CMB and SNIa data.}}

For simplification, we consider only the nearby GW sources, i.e. BNSs or NSBHs, at $z=0.1$. However, not all the binaries in this redshift can be used to constrain $H_0$. We select the sources with criteria of SNR$\ge 8$ and $\Delta d_{\rm L}/d_{\rm L}\le0.5$. The fraction of effective samples is given by $f_{H_0}=N_{\rm eff}/N_{\rm tot}$, where $N_{\rm eff}$ is the number of the effective samples that satisfy the selection criteria, and $N_{\rm tot}$ is the total simulated samples considered. We simulate $N_{\rm tot}=10^5$ samples at $z=0.1$ with random sky directions, inclination angles and polarization angles, and calculate the values of $f_{H_0}$ for various detector networks. The results are listed in Tables \ref{table2} and \ref{table3}. We find that for the 3G networks, nearly all the samples at $z=0.1$ satisfy the criteria, regardless of the detector types or network configurations.

%For each sample, the uncertainty of luminosity distance is calculated by the Fisher information matrix as above. We select the sources with criteria of SNR$\ge 8$ and $\Delta d_{\rm L}/d_{\rm L}\le0.5$. The faction of effective samples is given by $f_{H_0}=N_{\rm eff}/N_{\rm tot}$, where $N_{\rm eff}$ is the number of the effective sample, which satisfy the selection criteria, and $N_{\rm tot}$ is the total simulated samples, which is $10^5$ in our calculation.

%, and simulate $10^5$ samples at this redshift but with random sky-positions, inclination angles and polarization angles.

For a set of inspiral events with known redshift, the uncertainty of $H_0$ can be estimated by the following formula
\begin{equation}\label{delta-H0}
\frac{\Delta H_0}{H_0}=\left\{\sum_{k}\left(\frac{\Delta d_{\rm L}}{d_{\rm L}}\right)_k^{-2}\right\}^{-1/2},
\end{equation}
where $k=1,2,3,\cdot\cdot\cdot,N$, represents the $k$-th selected GW event. %with the parameter set of $\hat{\gamma}=(\theta_s,\phi_s,\iota,\psi_s)$ among the $N$ sources.
%, labels the event at $\hat{\gamma}_k$, where the vector $\hat{\gamma}$ stands for the angles
The quantity $(\Delta d_{\rm L}/d_{\rm L})_k$ is the uncertainty of distance of the $k$-th source. We have ignored in Eq. (\ref{delta-H0}) the photometric redshift errors and the possible errors generated by the peculiar velocities of the sources relative to the Hubble flow \cite{z-error}. When the number of the selected GW events is large, Eq. (\ref{delta-H0}) can be replaced by the following form
\begin{equation}
\frac{\Delta H_0}{H_0}=\frac{A_{H_0}}{\sqrt{N}}~.
\end{equation}
Note that, the total number of the effective GW events $N$ should be assumed in our calculation. The coefficient $A^{-2}_{H_0}$ is the average of $1/(\Delta d_{\rm L}/d_{\rm L}(\hat{\gamma}))^2$ over the angles $\hat{\gamma}=(\theta_s,\phi_s,\iota,\psi_s)$, that is
\begin{equation}
A^{-2}_{H_0}=\left\langle \frac{1}{(\Delta d_{\rm L}/d_{\rm L}(\hat{\gamma}))^2}\right\rangle_{\hat{\gamma}}.
\end{equation}
In order to evaluate $A_{H_0}$, we utilize all the effective samples, selected from $10^5$ random samples as mentioned above. The results of various detector networks are listed in Table \ref{table2} and Table \ref{table3} for BNSs and NSBHs respectively.

%marks the GW events.

%The result can be formally rewritten as follows
%\begin{equation}
%\frac{\Delta H_0}{H_0}=\frac{A_{H_0}}{\sqrt{N_{\rm eff}}}.
%\end{equation}
%Here, $N_{\rm eff}$ is the number of the effective samples, which satisfy the selection criteria. The fraction of effective samples is given by $f_{H_0}=N_{\rm eff}/N_{\rm tot}$, where $N_{\rm tot}$ is the total simulated samples, which is $10^5$ in our calculation. So, for a given detector network, the constraint on $H_0$ is quantified by two quantities $A_{H_0}$ and $f_{H_0}$. Based on $10^5$ random samples, we derive the their values for various detector networks, which are listed in Table \ref{table2} for BNSs and in Table \ref{table3} for NSBHs.

{For the BNSs, Table \ref{table2} shows that the 2CE and CE-ETD networks give the similar constrains: $\Delta H_0/H_0\simeq 1.3\%$ could be achieved if 40 sources are selected, which is comparable to the current accuracy. For the 2ETD network, 3-detector, or 4-detector networks, the constraint on $H_0$ becomes much more stringent, with $\Delta H_0/H_0$ around four times smaller. For these networks, a total of 10 selected sources can already achieve $\Delta H_0/H_0\simeq 0.6\%$, which is two times smaller than the best constraint on $H_0$ available at this writing. In comparison with BNSs, NSBHs always yield slightly worse constraint on $H_0$, expect for the CE-ETD network where the results are better. Note the best constraint of $H_0$ is mainly contributed by the GW sources with the best distance determination, i.e. the edge-on binaries, instead of the face-on ones.}

\subsection{Determination of deceleration parameter}

The deceleration parameter $q_0$ in cosmology is a dimensionless measure of the cosmic acceleration of the expansion of space in an FLRW universe. It is defined as $q_0\equiv -\left.\frac{\ddot{a}a}{\dot{a}^2}\right|_{t=t_0}$, where $a$ is the scale factor of the universe and the dots indicate derivatives by the proper time $t$ with $t_0$ representing the present time. A measurement of a negative $q_0$ indicates the accelerating expansion of the Universe at the present time. In the flat $\Lambda$CDM model, $q_0$ is directly related to the effective EoS of the universe $w$ by the relation $q_0=\frac{3}{2}(1+3w)$, or equivalently it is related to the matter density parameter $\Omega_{m}$ by the relation $q_0=\frac{3}{2}\Omega_{m}-1$. Latest observations give $\Omega_m=0.308\pm0.012$ \cite{planck}, which means $q_0=-0.538\pm0.018$, and strongly supports the accelerating expansion model of the Universe.

We investigate the potential constraint of $q_0$ by detected GW events at redshift $z<2$, through the observations of the 3G detector networks. For a flat $\Lambda$CDM model, $q_0$, or equivalently $\Omega_m$, is related to the luminosity distance $d_{\rm L}$ by \cite{weinberg},
\begin{equation}
d_{\rm L}(z)=\frac{1+z}{H_0}\int_0^z\frac{dz'}{\sqrt{\Omega_{m}(1+z')^3+(1-\Omega_{m})}},
\end{equation}
which depends on both $H_0$ and $\Omega_m$ (or, equivalently $q_0$). In this subsection, we assume that the Hubble constant $H_0$ is already well measured, either by other cosmological observations, such as the CMB observations \cite{planck}, or by the observations of the nearby GW events as discussed in the last section, where $H_0$ can be measured with unprecedented accuracy. Thus, assuming that the redshifts are known, the accuracy of measuring $q_0$ using these binary events depends solely on the uncertainties in measuring their luminosity distance $d_{\rm L}$. Specifically, the uncertainty of $q_0$ can be estimated by the formula
\begin{equation}\label{delta-q0}
\Delta q_0=\left\{\sum_{k}\left(\frac{\partial \ln d_{\rm L}/\partial q_0}{\delta d_{\rm L}/d_{\rm L}}\right)_k^{2}\right\}^{-1/2},
\end{equation}
where the values of $\partial \ln d_{\rm L}/\partial q_0$ are numerically calculated. For the high-$z$ GW events, distance measurements are subject to two kinds of uncertainties: the statistical error $\Delta d_{\rm L}/d_{\rm L}$ which can be estimated using a Fisher matrix as discussed above, and an additional error due to the effects of weak lensing \footnote{For biased estimation method, we also need to consider systematic errors.}. Following previous works in \cite{sathya2009,zhao-2011}, we assume the contribution to the distance error from weak lensing follows the relation  $\tilde{\Delta} d_{\rm L}/d_{\rm L}=0.05~z$, and that the total uncertainty of
$d_{\rm L}$ is taken to be $\delta d_{\rm L}/d_{\rm L}=\sqrt{(\Delta d_{\rm L}/d_{\rm L})^2+(\tilde{\Delta} d_{\rm L}/d_{\rm L})^2}$.

Similarly, the relation (\ref{delta-q0}) can be formally written as
\begin{equation}
\Delta q_0=\frac{A_{q_0}}{\sqrt{N}},
\end{equation}
where $N$ is the number of the effective GW events. The coefficient $A_{q_0}$ is given by
\begin{equation}\label{tt}
A^{-2}_{q_0}=\int_0^{z_{\max}} \left(\frac{\partial \ln d_{\rm L}(z)}{\partial q_0}\right)^2 f(z) \left\langle \frac{1}{(\delta d_{\rm L}/d_{\rm L}(\hat{\gamma},z))^2}\right\rangle_{\hat{\gamma}} dz,
\end{equation}
where $f(z)$ is the number distribution of GW sources over redshift $z$, that is normalized by the total number of events so that its integration over $z$ yields unity. In this paper, we assume that the GW sources are uniformly distributed in the comoving volume.

We evaluate numerically $A_{q_0}$ and the fraction of selected events $f_{q_0}$ by simulating $10^5$ random samples of BNSs and NSBHs with random sky directions, inclination angles and polarization angles. The samples are uniformly distributed in the comoving volume in the range $z<2$. The fraction of effective samples is also defined by $f_{q_0}=N_{\rm eff}/N_{\rm tot}$. The values of $A_{q_0}$ and $f_{q_0}$ quantify the measurement accuracy of $q_0$ by the detector network, and are listed in Tables \ref{table2} and \ref{table3} for BNSs and NSBHs respectively. Consistent with the previous results, for the 2CE network, we find that only a small fraction of binary coalescences satisfy the selection criteria with $f_{q_0}=3.98\%$ for BNSs and $6.29\%$ for NSBHs. For the CE-ETD network, we find that the values of $f_{q_0}$ are ten times lager. While for the rest of the networks studied, around $70\%$ or more sources can satisfy the selection criteria and are used to measure $q_0$. For these networks, $\Delta q_0\sim 0.02$ can be achieved if we select 100 BNSs or NSBHs with known redshift, which is already similar to the best accuracy available at this writing. Note the uncertainty of our estimation mainly depends on the number of the selected GW events, with accurate redshift information, that is much smaller than the total number of observable GW events in this redshift range ($\sim 10^7$ per year for BNSs) \cite{sathya2009}.

%{\r discussions.................The current Hubble constant is $H_0=67.8\pm 0.9~{\rm km~s}^{-1}{\rm~Mpc}^{-1}$ \cite{planck}, derived from the combination of cosmic microwave background radiation and other large-scale observations.......}

It has been argued that for high-$z$ binary coalescence GW events, the most promising method to measure their redshifts is to observe their GRB counterparts and afterglows \cite{gamma-ray}. On the other hand, GRBs are believed to be beamed: the $\gamma$ radiation is emitted in a narrow cone more or less perpendicular to the binary orbital plane, and the observed GRBs are nearly all beamed towards the Earth. For those face-on binaries with observed GRB counterparts, the sky direction ($\theta_s,\phi_s$), inclination angle $\iota$, and polarization angle $\psi_s$ can be determined precisely by the EM observation, therefore should be excluded in the Fisher matrix analysis.
%So, we should only consider the five-parameter Fisher matrix, and the uncertainty of $d_{\rm L}$ will be greatly reduced, with the expense of much less observable sources. In the previous works \cite{sathya2009,zhao-2011}, it is assumed that only $1,000$ GW events, less than $0.1\%$ of the total events, can be observed by ET. Here, we also consider this case based on $10^5$ face-on GW sources in the redshift range $z\in[0.1,2]$.
We repeat the same calculation for $10^5$ face-on GW sources in the redshift $z<2$ as above but adopting 5-parameter Fisher matrix, and recalculate $A'_{q_0}$ and $f'_{q_0}$ (listed in Tables \ref{table2} and \ref{table3}), which mimic $A_{q_0}$ and $f_{q_0}$ respectively. Note that, the quantity $\hat{\gamma}$ in this case represents $(\theta_s,\phi_s,\psi_s)$, instead of $(\theta_s,\phi_s,\iota,\psi_s)$.

We find that in this case, the uncertainty of distance become much smaller than that from the 9-parameter Fisher matrix analysis. All the simulated samples can satisfy the selection criteria, regardless of the network configurations. Due to larger SNRs of these face-on sources, the detector networks consisting of one or more CEs yield the best measurement accuracy of $q_0$. For a conservative estimation of 1000 observed BNSs or NSBHs as in previous works \cite{sathya2009,zhao-2011}, we have $\Delta q_0\sim 0.002$, which is 10 times smaller than the best measurement available presently. For other networks, the constraint becomes slightly less stringent, and the values of $\Delta q_0$ is $\sim 20\%$ larger for 2-detector or 3-detector networks, and $\sim 10\%$ larger for 4-detector networks. We should emphasize that, although the detector networks consisting of at least one ET-D give a slightly worse constraint on $q_0$ for the same number of the observed events, these networks yield a much better angular resolution for GW sources, which could be helpful to hunt for their EM counterparts, and greatly increase the total number of events with known redshift information.

\subsection{Determination of cosmic dark energy}

A possible explanation of cosmic acceleration could be the presence of dark energy, which has positive density but negative pressure (see reviews \cite{darkenergy,darkenergy-taskforce}). The key question is how well we will be able to differentiate between various dark energy models by measuring the dark energy EoS and its time evolution. A number of traditional EM methods have been used to constrain the EoS of dark energy, including using type Ia supernovae (SNIa), the temperature and polarization anisotropies of the CMB radiation, the baryon acoustic oscillations (BAO) peak in the distribution of galaxies, and weak gravitational lensing \cite{darkenergy,darkenergy-taskforce}. As mentioned above, GW standard siren provides a new method to probe the physics of dark energy, which have been partly discussed in the previous works \cite{sathya2009,zhao-2011} for a single ET detector. In this subsection, we extend the same analysis to the 3G detector networks.

Similar to \cite{zhao-2011}, we adopt a phenomenological form for EoS parameter $w$ as a function of redshift
\begin{equation}
w(z)\equiv\frac{p_{\rm de}}{\rho_{\rm de}}=w_0+w_a\frac{z}{1+z},
\end{equation}
where $p_{\rm de}$ and $\rho_{\rm de}$ are the pressure and energy density of dark energy respectively. This is the widely adopted Chevalier-Polarski-Linder form in the literature \cite{CPL,darkenergy-taskforce}. In this form, $w_0$ is the present EoS and $w_a$ quantifies its evolution with redshift. In the $\Lambda$CDM model, dark energy is described by the cosmological constant, which has $w_0=-1$ and $w_a=0$. We consider a general FLRW universe, in which the luminosity distance of astrophysical sources as a function of redshift $z$ is given by \cite{weinberg}
\begin{eqnarray}
d_L(z)=(1+z)\left\{
\begin{array}{c}
      {{|k|}^{-1/2}}\sin\left[{|k|}^{1/2}\int^z_0\frac{dz'}{H(z')}\right]~~~(\Omega_k<0),\\
    \int^z_0\frac{dz'}{H(z')}~~~~~~~~~~~~~~~~~~~~~~~~~~~~(\Omega_k=0), \label{dl}\\
      {{|k|}^{-1/2}}\sinh\left[{|k|}^{1/2}\int^z_0\frac{dz'}{H(z')}\right]~~~(\Omega_k>0),
 \end{array}
 \right.
 \end{eqnarray}
where $|k|^{1/2}\equiv H_0\sqrt{|\Omega_k|}$ and $\Omega_k$ is the contribution of spatial curvature to the critical density. The Hubble parameter $H(z)$ is given by
\begin{equation}
H(z)=H_0[\Omega_m(1+z)^3+\Omega_k(1+z)^2+(1-\Omega_m-\Omega_k)E(z)]^{1/2},
\end{equation}
where the function $E(z)$ is defined as
\begin{equation}
E(z)=(1+z)^{3(1+w_0+w_a)}e^{-3w_az/(1+z)}.
\end{equation}

From the expression of $d_{\rm L}$, it seems possible that one can constrain the full parameter set $(H_0,\Omega_m,\Omega_k,w_0,w_a)$ together by the GW data alone, as long as the redshifts of GW sources are known. Unfortunately, in the previous work \cite{zhao-2011}, we found this globe constraints cannot be realized, due to the strong degeneracy between the background parameters $(H_0,\Omega_m,\Omega_k)$ and the dark energy parameters $(w_0,w_a)$. The same problem also happens in other methods for dark energy detection (e.g., SNIa and BAO methods). A general way to break this degeneracy is to combine the result with the CMB data, which are sensitive to the background parameters $(H_0,\Omega_m,\Omega_k)$, and provide the necessary complementarity to the GW data. It has also been discovered in \cite{zhao-2011} that, taking the CMB observation as a prior is nearly equivalent to treat the parameters $(H_0,\Omega_m,\Omega_k)$ as known in data analysis. Thus, we use the GW data to constrain the parameters $(w_0,w_a)$ only.

In order to estimate the errors of these parameters, we study a Fisher matrix $F_{ij}$ for a collection of inspiral events, which is given by
\begin{equation}\label{ss}
F_{ij}=\sum_{k}\frac{\left({\partial \ln d_{\rm L}(k)}/{\partial p_i}\right)\left({\partial \ln d_{\rm L}(k)}/{\partial p_j}\right)}{\left(\delta d_{\rm L}/d_{\rm L}(k)\right)^2},
\end{equation}
where the indices $i$ and $j$ run from 1 to 2, for the two free parameters ($w_0$, $w_a$). The index $k=1,2,\cdot\cdot\cdot,N$, labels the event at $(z_k,\hat{\gamma}_k)$ among the total $N$ sources. The value of $\delta d_{\rm L}/d_{\rm L}$ includes both the statistical error and the weak-lensing error. The uncertainties of dark energy parameters are given by $\Delta w_0=(F^{-1})_{11}$ and $\Delta w_a=(F^{-1})_{22}$. To estimate and compare the goodness of constraints from the observational data sets, we calculate the figure of merit (FoM) \cite{darkenergy-taskforce}, for each data sets, which is proportional to the inverse area of the error ellipse in the $w_0$-$w_a$ plane,
\begin{equation}\label{fom}
{\rm FoM}=\left[{\rm Det} ~C(w_0,w_a)\right]^{-1/2},
\end{equation}
where $C(w_0,w_a)$ is the covariance matrix of $w_0$ and $w_a$. Larger FoM means stronger constraint on the parameters since it corresponds to a smaller error ellipse. Similar to the discussion of $q_0$ constraint above, we consider only the face-on sources with known redshift and calculate their distance uncertainties with the 5-parameter Fisher matrix. We adopt a fiducial cosmological model with the values of parameters given by Planck Collaboration in \cite{planck}.

When the number of events is large, the sum over events in Eq. (\ref{ss}) can be replaced by the following integral \cite{zhao-2011},
\begin{equation}
F_{ij}=N \times \int_0^{z_{\max}} \frac{\partial \ln d_{\rm L}(z)}{\partial p_i} \frac{\partial \ln d_{\rm L}(z)}{\partial p_j} f(z) \left\langle \frac{1}{(\delta d_{\rm L}/d_{\rm L}(\hat{\gamma},z))^2}\right\rangle_{\hat{\gamma}} dz.
\end{equation}
The numerical calculation, based on $10^5$ random samples, gives the following results
\begin{equation}
\Delta w_0=\frac{A_{w_0}}{\sqrt{N}},~~~~\Delta w_a=\frac{A_{w_a}}{\sqrt{N}},~~~{\rm FoM}=A_{\rm FoM} N.
\end{equation}
The values of the coefficients $A_{w_0}$, $A_{w_a}$ and $A_{\rm FoM}$ for various detector networks are listed in Tables \ref{table2} and
\ref{table3}.

In Fig.~\ref{fe1}, we plot the two-dimensional uncertainty contours of various detector networks for demonstration.
For a conservative estimation with 1000 selected BNSs, for the 2CE network, we obtain,
\begin{equation}
\Delta w_0=0.032,~~~\Delta w_a=0.20,~~~{\rm FoM}=517.
\end{equation}
We find that the constraints of $w_0$ and $w_a$ are two times more stringent than the results for an single ET detector, derived in previous work \cite{zhao-2011}. For the 3CE network, the results are similar, with $\Delta w_0=0.030$, $\Delta w_a=0.19$ and ${\rm FoM}=566$. However, for the network consisting of at least one ET-D detector, the uncertainties are slightly larger, due to smaller SNRs. For instance, for the 2ETD network, we obtain $\Delta w_0=0.043$, $\Delta w_a=0.26$ and ${\rm FoM}=295$, which become $\Delta w_0=0.036$, $\Delta w_a=0.22$ and ${\rm FoM}=420$ for the 4ETD network.

We compare the capability of the 3G detectors in constraining the EoS of the dark energy with that of traditional EM methods, following the same analyses as in \cite{zhao-2011} for SNIa method and BAO method. For the future SNIa survey, we considered the SNAP (Supernova/Acceleration Probe) project \cite{snap,darkenergy-taskforce}, which considered 300 low redshift supernovae, uniformly distributed over $z\in(0.03,0.08)$, and 2000 high redshift supernovae in the range $z\in(0.1,1.7)$. For the potential BAO observation, we considered the finial JDEM (Joint Dark Energy Mission) project \cite{jdem,darkenergy-taskforce}, which is expected to survey the galaxies of $10,000$ deg$^2$ in the redshift range $z\in(0.5,2)$. The corresponding two-dimensional uncertainty contours of these two projects are also plotted in Fig.~\ref{fe1}, from which we find that with the 2CE 3G GW detector network, the constraints of dark energy parameters are significantly more stringent than the traditional SNIa and BAO methods, using conservatively only 1000 GW events.

%For ET-B alone, in the previous work \cite{zhao-2011}, we derived that $A_{w_0}=2.024$, $A_{w_a}=12.27$, $z_p=0.188$, $A_{w_p}=0.601$, which follow that for 1000 GW events, the uncertainties of dark energy parameters are
%\begin{equation}
%\Delta w_0=0.064,~~~\Delta w_a=0.388,~~~z_p=0.188,~~~\Delta w_p=0.019.
%\end{equation}

%For the case with negligible instrumental error $\Delta d_{\rm L}/d_{\rm L}$, i.e. the uncertainty of luminosity distance is completely contributed by the weak lensing effect, we obtain that that $A_{w_0}=0.843$, $A_{w_a}=5.352$, $z_p=0.176$, $A_{w_p}=0.259$, which follow that for 1000 GW events, the uncertainties of dark energy parameters are
%\begin{equation}
%\Delta w_0=0.027,~~~\Delta w_a=0.169,~~~z_p=0.176,~~~\Delta w_p=0.008.
%\end{equation}

%************************************************************************   table 1  ******      table 1  *****************************************************************************************
\begin{table}
\caption{The constraints of Hubble constant $H_0$ and deceleration parameter $q_0$, and EoS of dark energy derived from $10^5$ randomly distributed BNS samples.}
\begin{center}
\label{table2}
\begin{tabular}{|c| c| c| c| c| c| c| c|}
    \hline
     & ~~~~2CE~~~~ & ~~~~CE-ETD~~~~ & ~~~~2ETD~~~~ &~~~~3CE~~~~ &~~~~3ETD~~~~&~~~~3cETD~~~~&~~~~4ETD~~~~\\
         \hline
    $A_{H_0}$ & 0.077 & 0.082 & 0.029 & 0.015 &  0.022 & 0.022  & 0.019 \\
         \hline
    $f_{H_0}$ & 81.4\%& 90.6\%  & 97.8\% & 99.7\% &98.8\% & 99.1\%&99.5\% \\
         \hline
         \hline
    $A_{q_0}$ & 0.335 & 0.337 & 0.262 & 0.160 &  0.216 & 0.216  & 0.193  \\
         \hline
    $f_{q_0}$ & 3.98\%& 38.0\%  & 66.3\% & 77.2\% &75.6\% & 77.6\%& 80.5\%\\
         \hline
         \hline
    $A'_{q_0}$ & 0.071 & 0.088 & 0.092 & 0.068 &  0.084 & 0.082  & 0.078 \\
         \hline
    $f'_{q_0}$ & 100\%& 100\%  & 100\% & 100\% & 100\% & 100\% & 100\% \\
         \hline
         \hline
    $A_{w_0}$ &  1.007 & 1.506 & 1.350 & 0.959 & 1.216 & 1.197 & 1.124 \\
         \hline
    $A_{w_a}$ &  6.326 & 8.963 & 8.380 & 6.040 & 7.572 & 7.457 & 7.020 \\
         \hline
    $A_{\rm FoM}$ &  0.517 & 0.280 & 0.295 & 0.566 & 0.361 & 0.372 & 0.420 \\
         \hline
  %  $z_p$ &     0.179  & 0.193 & 0.182 & 0.178 & 0.181 & 0.181 & 0.180 \\
  %       \hline
  %  $A_{w_p}$ & 0.306  & 0.398 & 0.404 & 0.292 & 0.366 & 0.360 & 0.339 \\
  %       \hline
\end{tabular}
\end{center}
\end{table}
%************************************************************************   table 1  ******      table 1 *****************************************************************************************

%************************************************************************   table 1  ******      table 1  *****************************************************************************************
\begin{table}
\caption{Same as Table \ref{table2}, but here the NSBHs, instead of BNSs, are considered.}
\begin{center}
\label{table3}
\begin{tabular}{|c| c| c| c| c| c| c| c|}
    \hline
     & ~~~~2CE~~~~ & ~~~~CE-ETD~~~~ & ~~~~2ETD~~~~ &~~~~3CE~~~~ &~~~~3ETD~~~~&~~~~3cETD~~~~&~~~~4ETD~~~~\\
         \hline
    $A_{H_0}$ & 0.092 & 0.056 & 0.048 & 0.025 & 0.037  & 0.037  & 0.032 \\
         \hline
    $f_{H_0}$ & 97.4\%& 100\%  & 100\% & 100\% & 100\% & 100\% & 100\% \\
         \hline
         \hline
    $A_{q_0}$ & 0.362 & 0.239 & 0.213 & 0.143 &  0.177 & 0.175  & 0.158 \\
         \hline
    $f_{q_0}$ & 6.29\%& 61.3\%  & 76.4\% & 84.9\% & 83.7\% & 85.0\%& 87.5\%\\
         \hline
         \hline
    $A'_{q_0}$ & 0.066 & 0.070 & 0.080 & 0.065 &  0.074 & 0.073  & 0.071 \\
         \hline
    $f'_{q_0}$ & 100\%& 100\%  & 100\% & 100\% & 100\%&100\% &100\% \\
         \hline
         \hline
    $A_{w_0}$ & 1.012 & 1.104 & 1.334 & 0.965 & 1.211 & 1.194 & 1.126 \\
         \hline
    $A_{w_a}$ & 6.259 & 6.760 & 8.015 & 6.008 & 7.341 & 7.243 & 6.875 \\
         \hline
    $A_{\rm FoM}$ &  0.549 & 0.476 & 0.347 & 0.590 & 0.410 & 0.421 & 0.463 \\
         \hline
  %  $z_p$ &  0.183    & 0.186 & 0.191 & 0.182 & 0.188 & 0.188 & 0.187 \\
  %       \hline
  %  $A_{w_p}$ & 0.291 & 0.310 & 0.359 & 0.282 & 0.332 & 0.328 & 0.314 \\
  %       \hline
\end{tabular}
\end{center}
\end{table}
%************************************************************************   table 1  ******      table 1 *****************************************************************************************

\begin{figure}
\begin{center}
\centerline{\includegraphics[width=17cm]{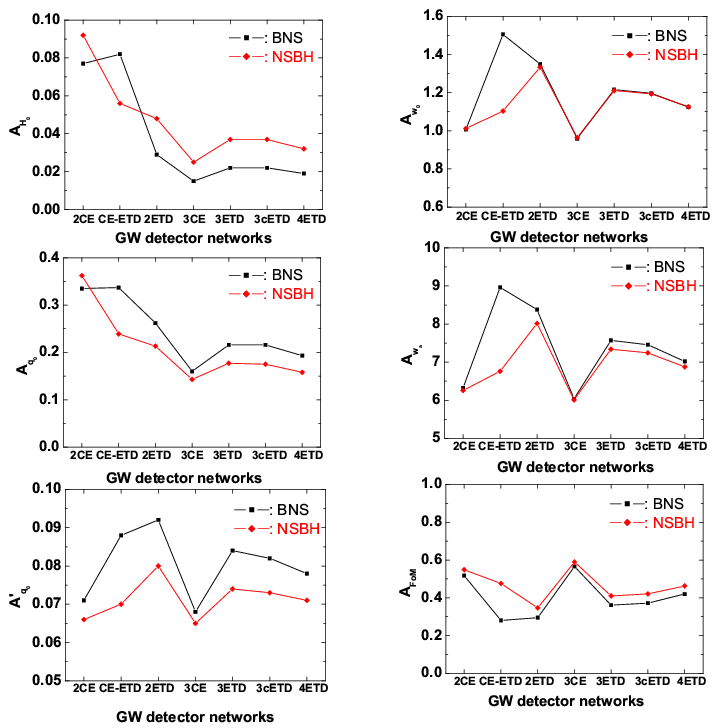}}
\end{center}\caption{The values of $A_{H_0}$ (upper left), $A_{q_0}$ (middle left), $A'_{q_0}$ (lower left), $A_{w_0}$ (upper right),
$A_{w_a}$ (middle right), $A_{\rm FoM}$ (lower right) for various GW detector networks. In each panel, the black dots denote the results derived from $10^5$ randomly distributed BNS samples, and the red dots denote the results derived from $10^5$ randomly distributed NSBH samples.}\label{cos2}
\end{figure}

\begin{figure}
\begin{center}
\centerline{\includegraphics[width=15cm,height=10cm]{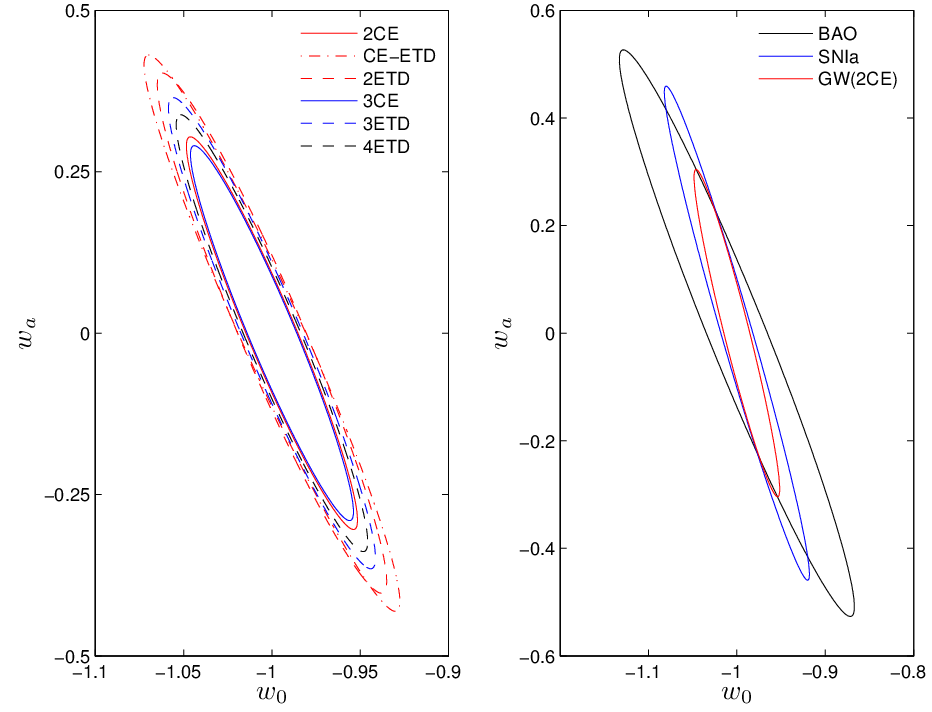}}
\end{center}\caption{The two-dimensional uncertainty contours of the dark energy parameters $w_0$ and $w_a$ for various detection methods. Note that, in the case of GW detector networks, we assumed that $1000$ face-on BNS sources are observed.}\label{fe1}
\end{figure}

\section{Conclusions}
\label{sec:Concl}

The detection of GW sources opens the possibility of using a brand new GW window to further our understanding of fundamental physics, cosmology and astrophysics. The coalescing compact binaries are the prime sources of ground-based detectors. The localization of them, including the angular resolution and the distance determination, is one of the key tasks for the GW observations. As the standard sirens, the luminosity distance of these binary coalescence sources can be precisely measured by GW observations without the aid of any cosmic distance ladder. Meanwhile, the accurate localization of GW sources is helpful for the identification of their host galaxies or the follow-up observations on the EM counterparts, which enables independent measurements of their redshifts. Thus, the distance-redshift relation can be used to measure the expansion history of the Universe, including the determination of the Hubble constant, the acceleration of the Universe, and measuring the EoS of dark energy.

In this article, we give a comprehensive discussion on the localization of GW events by 3G GW detectors, which aim to significantly increase the detection band toward frequencies as low as a few Hz. Taking into account the rotation of the Earth, a single detector can be effectively treated as a detector network with long baselines. Thus, the localization of binary coalescence sources with small masses could be greatly improved, as long as the contribution of GW signal at sub-10 Hz range is significant. In this article, we employ the standard Fisher information matrix to investigate the localization capabilities of 3G detectors, including a single detector, and the network consisting of up to four detectors with different configurations. Focusing on the ET and CE detectors, we first investigate the influence of the Earth's rotation on the localization of GW sources. We find that, for an individual 3G detector (in particular for ET), the localization of coalescing BNSs and low-mass NSBHs, can be significantly improved due to the periodic self-rotation of the Earth. This effect is much more significant for the edge-on binaries, than the face-on ones. In addition, we find that a slight reduction of the detector noise at low-frequency range, in particular in the range $f\in(1,10)$ Hz, can significantly improve the localizations of GW sources by an individual 3G detector. For ET and an ideal detector with flat low-frequency noise level, the dependence of $\Delta\Omega_s$ and $\Delta d_{\rm L}/d_{\rm L}$ on the low-frequency cutoff can be approximated by a broken power-law formula.

As the main task of this article, we investigate the localization capabilities of 3G experiments as a function of redshift for BNS and NSBHs. For a network of two CE detectors, nearby BNSs or NSBHs ($z\lesssim 0.1$) can be localized with reasonable accuracy of $\Delta\Omega_s\sim 10~$deg$^2$ and $\Delta d_{\rm L}/d_{\rm L}<1$. However, if at least one ET-D detector is included in the 2-detector networks, the localization of GW sources can be greatly improved, due to the special configuration of ET experiment. For instance, for both BNSs and NSBHs at $z=0.1$, we have $\Delta\Omega_s\sim 1~$deg$^2$ and $\Delta d_{\rm L}/d_{\rm L}\sim ~0.1$. If considering a promising network consisting of four ET-D detectors in the Europe, US, China and Australia, the angular resolution of the BNSs and NSBHs at $z=0.1$ could reach the accuracy of $\Delta\Omega_s\sim 0.1~$deg$^2$ and $\Delta d_{\rm L}/d_{\rm L} \sim 0.003$, which might be sufficient to directly identify their host galaxies \cite{Loveday}. In addition, for 4-detector network, the accurate localization of GW sources can be extended to higher redshifts. For BNSs and NSBHs at $z=2$, we have $\Delta\Omega_s\sim 10~$deg$^2$, which is comparable to the field of view of some EM telescopes (such as Pan-STARRS and SkyMapper telescope). Thus, confident identifications of their EM counterparts become possible, and these sources can be extremely important for high-$z$ multi-messenger astronomy including cosmology. Our result also show that, for an individual 3G detector and the detector networks, there is a pronounced anti-correlation between the angular resolution and the distance accuracy. For GW sources at a fixed redshift, the face-on ones yield a better angular resolution, while the edge-on ones have a better distance determination.

We discuss the implication of our result to using GW sources as an independent tool to probe the cosmological model of our Universe.  In particular, we investigate how detections of GW sources by 3G detectors can help constraining the Hubble constant $H_0$, the deceleration parameter $q_0$, and the EoS of cosmic dark energy. We conclude that the 2-detector network 2ETD, 3-detector, and 4-detector networks could help set stringent constraint on $H_0$ using BNSs or NSBH at low-redshift. Even if only 10 BNSs or NSBHs with known redshift, selected with criteria of SNR$\ge 8$ and $\Delta d_{\rm L}/d_{\rm L}\le 0.5$, can be observed by one of these detector networks, the accuracy of $H_0$ can be achieved at $0.9\%$ or less, which is two times smaller than the currently best accuracy. %If the observed values of $\Delta d_{\rm L}/d_{\rm L}$ are larger than those evaluated from Fisher matrix analysis by a factor 2 or 5, the uncertainty of $H_0$ will increase by a same factor.
In addition, we find that the edge-on sources, instead of the face-on sources, dominate the constraint of $H_0$ due to better $\Delta d_{\rm L}/d_{\rm L}$ for edge-on ones. For constraining $q_0$, we find that the capabilities of various 3G detector networks are similar. Considering the face-on GW events at the redshift range $z<2$, for which the redshifts are possible to be measured from their EM counterparts, we find that if 1000 events can be selected, $\Delta q_0\sim 0.002$ could be achieved, which is around 10 times better than the currently best values. These 1000 face-on events can also constrain the EoS of cosmic dark energy with accuracies $\Delta w_0\sim 0.03$ and $\Delta w_a\sim 0.2$, which are more stringent than the projected accuracies for the future JDEM BAO project and the SNAP SNIa observations. These results show that the cosmography with 3G detector networks is extremely promising, which provides an independent method to probe our Universe up to the high redshift range.

~

~

~

\begin{acknowledgments}

 We appreciate the helpful discussions with Man Leong Chan, Ik Siong Heng, Martin Hendry, Yiming Hu, Christopher Messenger, David Blair, Chunnong Zhao and Li Ju. WZ is supported by NSFC Grants Nos. 11773028, 11603020, 11633001, 11173021, 11322324, 11653002 and 11421303, the project of Knowledge Innovation Program of Chinese Academy of Science, the Fundamental Research Funds for the Central Universities and the Strategic Priority Research Program of the Chinese Academy of Sciences Grant No. XDB23010200. LW's research is supported in part by Australian Research Council.
\end{acknowledgments}

\end{document}